# Controversies and Progress in the Problem of Dynamical Fluctuation –Electromagnetic Interactions

A.A. Kyasov and G.V. Dedkov[1]

Nanoscale Physics Group, Kabardino-Balkarian State University, Nalchik, 360004, Russian Federation

A review is given of the key works relevant to the problem of dynamical fluctuation-electromagnetic interaction---a problem of topical interest over a period of several decades. We discuss controversies and present day progress in this area considering two basic geometrical configurations: a small particle moving above a thick plate, and two parallel plates in relative motion, separated by vacuum gap.

## 1.Introduction

Soon after its creation by Rytov [1], the theory of electromagnetic fluctuations has been successfully applied by Lifshitz [2] to the problem of van der Waals interaction between two resting dielectric plates at thermal equilibrium, and researchers have got a powerful method making it possible to study fluctuation-electromagnetic interaction between extended bodies and small particles with the bodies of different geometry. In literature this method is sometimes called fluctuation electrodynamics. Its essence is the solution of Maxwell's equations with the additional fluctuating sources, such as the discrete fluctuating multipole moments of a small particle or continuous fluctuating densities of electric/magnetic currents of matter. In addition, all physical quantities that characterize fluctuation-electromagnetic interaction (the stress tensor and Pointing's vector, etc.) are expressed through the integrals of quadratic forms related to spontaneous fluctuating sources. Quantum and statistical averaging of these forms is made using the fluctuation-dissipative relationships.

In some cases, rather than solve Maxwell's equations with fluctuating sources, is it more convenient to use the retarded Green function $D^R{}_{lm}(\omega,\mathbf{r},\mathbf{r}')$ which is obtained from the equation

$$\left(rot_{ik}rot_{kl} - \frac{\omega^2}{c^2}\varepsilon(\omega)\delta_{il}\right)D^R{}_{lm}(\omega,\mathbf{r},\mathbf{r}') = -4\pi\hbar\delta_{im}\delta(\mathbf{r}-\mathbf{r}') \;, \tag{1}$$

---
[1] Corresponding author e-mail: gv_dedkov@mail.ru

subjected to the necessary boundary conditions on the surfaces of extended bodies [3]. The spectral densities of correlators from various components of equilibrium electromagnetic field in the medium are expressed through the retarded Green function [3].

Therefore, by the beginning of 1960s, the major prerequisites were obtained for studying dynamical problems of fluctuation-electromagnetic interactions at least in two important configurations: a small particle above a thick plate (configuration 1) and two plates (configuration 2). Nevertheless, a rather long period of time has passed prior to appearing first works on this problem. Can there be a transverse dissipative (frictional) force between the bodies in relative motion ? To date, there are serious yet unresolved controversies in this problem [4]. Another dynamical situation that is recently intensively investigated both theoretically and experimentally refers to the so-called dynamical Casimir effect---the generation of photons out of the quantum vacuum driven by a moving mirror (Moore (1970), Dalvit et. al. (2010)). This problem is out of scope of this paper, but we briefly touch it in Sec. 4.2 (see the references therein). As well, we do not discuss (in detail) the problem of radiative heat exchange, which is caused by fluctuation-electromagnetic interaction between the neutral bodies with different temperatures, though the corresponding formulas for the heating rate of the moving particle are also given (see Secs. 3.4 and 3.5).

Since our first works on dynamic fluctuation-electromagnetic interactions [5], we have attacked this problem using a unitary theoretical basis of fluctuation electrodynamics, calculating the conservative –dissipative forces and heating rate between the uniformly moving bodies with different temperature. From the very beginning, our basic configuration was assumed to be configuration 1, for which it is possible to formulate and solve the problem in an unambiguous way. Using the exact solution obtained in configuration 1, we have formulated the correspondence rules between configurations 1 and 2. This allows us to find several important results in configuration 2, as well, though with some loss of generality. The aim of this work is to give a comparative review of our results with those obtained by other authors. We draw attention to the fact that a popular recipe in which the frequency in arguments of the particle polarizability or the reflection coefficient of a plane surface are replaced by the Doppler-shifted frequencies may result in catastrophic errors when solving the dynamical problem, especially in the case of relativistic velocities. Such errors are precisely these that were made by many authors.

The structure of this work is the following. In Sec.2 we present the early historical perspective relevant to the dynamical fluctuation-electromagnetic interaction, where we critically analyze the most known works of different authors between 1978 and 1996. Section 3 is devoted to the key works during the last two decades. Particularly, we recall exact solutions to the dynamical problem 1 and formulate the correspondence rules between configurations 1 and 2.

Based on these correspondence rules, we perform dynamical and nonequilibrium generalization of Lifshitz's theory in configuration 2. The remained unresolved problems relevant to the relativistic problem in configuration 2 are discussed. Section 4 contains a brief review of experimental studies, and Section 5---concluding remarks. Throughout the paper (with minor exceptions indicated) we use Gaussian units and conventional designations for $k_B, \hbar, c$ – the Boltzmann's and Planck's constants, and the speed of light in vacuum.

## 2. Genesis of the problem and historical perspective

### 2.1 Teodorovich (1978)

To our knowledge, the first attempt to find the contribution of van der Waals interactions to frictional force between two plates in relative motion separated by a vacuum gap was made by Teodorovich [6] (Fig. 1). In this work, the mean value of the Coulomb force between the fluctuating charges of the first and second plate is given by

$$\mathbf{F}_{12} = -\int d\mathbf{r}_1 d\mathbf{r}_2 \nabla_1 \left( \frac{1}{|\mathbf{r}_1 - \mathbf{r}_2|} \right) \langle \rho(1)\rho(2) \rangle \tag{2}$$

where $\rho(1) = \rho(\mathbf{r}_1, t)$ and $\rho(2) = \rho(\mathbf{r}_2, t)$ are the charge density operators in the moving plates 1 and 2, and angular brackets denote statistical averaging. To calculate the correlator $\langle \rho(1)\rho(2) \rangle$, Teodorovich finally solves the Poisson's equation of the form

$$\varepsilon(\omega, \mathbf{q}) \left( \frac{\partial^2}{\partial z_1^2} - q^2 \right) V_s(\mathbf{q}, z_1, z_2, \omega) = -4\pi \delta(z_1 - z_2) \tag{3}$$

where $\mathbf{q} = (k_x, k_y)$. In addition, the dielectric permittivity in Eq. (3) is given by

$$\varepsilon(\omega, \mathbf{q}) = \begin{cases} \varepsilon_1(\omega + 0.5\mathbf{qV}), & z > l/2 \\ \varepsilon_2(\omega - 0.5\mathbf{qV}), & z < l/2 \end{cases}$$

Therefore, direct solution of Eq. (1) or Maxwell's equations with fluctuating sources involving the necessary boundary conditions on moving plates at $z = \pm l/2$ was not carried out. Instead, the author used a heuristic replacement of the frequency $\omega$ by the Doppler-shifted frequencies $\omega \pm 0.5\mathbf{qV}$ in the expressions for the dielectric permittivities $\varepsilon_{1,2}(\omega)$ of the plates.

As a result, the expression for the tangential frictional force at the temperature $T = 0$ of the plates was obtained in the first nonvanishing order of velocity expansion. It is given by

$$\mathbf{F} = \frac{\hbar S \mathbf{V}}{128\pi^2 l^4} \int_0^\infty dx x^3 \left[ \frac{\varepsilon_{01} + 1}{\varepsilon_{01} - 1} \cdot \frac{\varepsilon_{02} + 1}{\varepsilon_{02} - 1} e^x - 1 \right]^{-1} \tag{4}$$

where $S$ is the surface area of vacuum contact, and $\varepsilon_{01}$, $\varepsilon_{02}$ are the static dielectric permittivities. Eq. (4) erroneously predicts the linear-velocity friction force at $T=0$. In addition, the dependence (4) on the static dielectric permittivities is inconsistent with the dissipative nature of friction force.

The formal transition $\varepsilon(\omega) \to \varepsilon(\omega + \mathbf{k}\mathbf{V})$ or $\alpha(\omega) \to \alpha(\omega + \mathbf{k}\mathbf{V})$ (the latter being the dielectric polarizability of the moving particle) from static to dynamic problems 1 and 2 was thereafter frequently used by many authors, who tried to generalize Lifshitz's theory.

## 2.2 Ferrell and Ritchie (1980)

Ferrell and Ritchie were the first who calculated dynamical van der Waals interaction potential between the moving ground state atom and a metal surface at $T=0$ [7] (Fig. 2). Using the second-order perturbation theory, the authors have firstly derived the expression (Eq.(8), Ref. 7) for the attractive van der Waals atom-surface energy involving the static case (hereafter in this Section the atomic units $e = \hbar = m = 1$ are used):

$$U(z_0) = -\frac{2}{3}i\sum_n (1-\delta_{1n}) \int_{-\infty}^{+\infty} \frac{d\omega}{2\pi} \Delta(\omega) \int_0^\infty dk\, k^2 \exp(-2k z_0) \frac{|\langle d_{1n}\rangle|^2}{(\omega - \omega_{1n})} \tag{5}$$

where $z_0$ is the atom-surface separation distance, $k = \sqrt{k_x^2 + k_y^2}$, $|\langle d\rangle_{1n}|^2 = 3|\langle 1|d_x|n\rangle|^2$ is the matrix element of the electron dipole moment operator taken using the ground state and the n-th excited state, $\omega_{n1} = \omega_n - \omega_1$, $\Delta(\omega) = (\varepsilon(\omega) - 1)/(\varepsilon(\omega) + 1)$. Eq.(5) is equivalent to the earlier result of Mavroyannis [8]

$$U(z_0) = -\frac{1}{6\pi z_0^3} \sum_n |\langle d\rangle_{1n}|^2 (1-\delta_{n1}) \int_0^\infty d\omega \frac{\omega_{n1}}{\omega_{n1}^2 + \omega^2} \Delta(i\omega) \tag{6}$$

Using (6), the authors have found the interaction energy in the dynamical case replacing $\omega$ by $\omega - kV\cos\theta$ and integrating over the polar angle $0 \le \theta \le 2\pi$ (see Eq.(14) in Ref. 7). This is equivalent to the transformation $\alpha(\omega) \to \alpha(\omega - k_x V)$ and subsequent integration over components of the two-dimensional wave vector $-\infty < k_x, k_y < \infty$. Applying a single oscillator model of atomic polarizability, the corresponding expression takes the form

$$U(z_0, V) = -\frac{2\alpha(0)\omega_0^2}{\pi^2} \int_0^\infty d\omega \int_0^\infty dk_x \int_0^\infty dk_y\, k \exp(-2k z_0) \Delta(i\omega) \frac{(\omega_0^2 + \omega^2 - k_x^2 V^2)}{(\omega_0^2 + \omega^2 - k_x^2 V^2)^2 + 4\omega^2 k_x^2 V^2} \tag{7}$$

Formula (7) is very similar to our recent result (Eq. (8) in [9]), that was obtained in the scope of consistent fluctuation-electromagnetic theory (see Section 3):

$$U(z_0,V) = -\frac{2\alpha(0)\omega_0^2}{\pi^2}\int_0^\infty d\omega \int_0^\infty dk_x \int_0^\infty dk_y\, k\exp(-2kz_0)\frac{\Delta(i\omega)(\omega_0^2+\omega^2-k_x^2V^2)}{(\omega_0^2+\omega^2-k_x^2V^2)^2+4\omega^2 k_x^2 V^2} + \quad (8)$$
$$+\Delta U(z_0,V)$$

$$\Delta U(z_0,V) = \frac{2}{\pi^2}\int_0^\infty dk_x \int_0^\infty dk_y\, k\exp(-2kz_0)\int_0^{k_x V} d\omega\, \Delta'(\omega)\alpha''(\omega-k_x V) \quad (9)$$

The only difference between (7) and (8) is the presence of the additional term $\Delta U(z_0,V)$. At small velocities of an atom, $V \ll \omega_0 z_0$, $\Delta U(z_0,V) \to 0$, Eqs. (7) and (8) lead to the same result

$$U(z_0,V) = -\frac{\alpha(0)}{8z_0^3}\frac{\omega_0 \omega_s}{\omega_0+\omega_s}\left(1+\frac{3V^2}{2z_0^2}\frac{1}{(\omega_0+\omega_s)^2}\right) \quad (10)$$

where $\omega_s = \omega_p/\sqrt{2}$ and $\omega_p$ is the plasma frequency of metal plate, but the difference between (7) and (8) becomes crucially important if the velocity increases [9,10]. Therefore, the heuristic transition from static to dynamic van der Waals atom-surface interaction using the Doppler-shifted frequency in the polarizability of the moving particle is generally incorrect. Nevertheless, the work [7] has not presently lost its practical importance owing to conceptual clarity. Eq. (10) is considered as the reference result and has been subsequently reproduced by several authors using different calculation methods [9, 11-13].

### 2.3 Mahanty, Schaich & Harris (1980,1981)

As in the aforementioned works, Mahanty, Schaich and Harris [14,15] have examined the special case of nonretarded van der Waals interactions at $T=0$ [14], and both at $T=0$ and $T>0$ [15]. Apart from the atom-surface interaction forces, Mahanty has calculated the velocity-dependent force between the two molecules in relative motion, as well. He used the known (in static case) expression for the distance-dependent free energy of two interacting bodies [16]

$$U(z_0) = \frac{\hbar}{2\pi}\int_0^\infty d\xi \ln D(i\xi) \quad (11)$$

where $D(\omega)$ is the dispersion equation for the normal electromagnetic modes of the system. The moving particle was constrained to move with constant velocity along the given trajectory, and the time Fourier transforms of the induced dipole moments were taken at the Doppler-shifted frequencies. Particularly, the tangential force on a particle moving parallel to the surface was obtained as the derivative of free energy with respect to the lateral coordinate. The lateral dissipative force was then given by $F \propto V/z_0^5$. The general shortcoming of this approach is the lack of strict solution to the Maxwell's equations with moving dipole sources. This leads to an incorrect behavior of the atom-surface energy, because the corresponding expression (see Eq. (20)) in [14]) contains the Doppler-shifted frequency both in the argument of the surface response function and the polarizability. Particularly, this makes it impossible to obtain Eq. (10). Correction term (9) is also absent. Additional shortcoming is that the dissipative force is calculated as the derivative of the conservative interaction energy.

Contrary to the works under discussion, Schaich and Harris [15] have used the formally exact expression for the friction coefficient

$$\gamma = (k_B T)^{-1} \text{Re} \int_0^\infty dt \left\langle \hat{\mathbf{F}}(t)\hat{\mathbf{F}}(0) \right\rangle_R \tag{12}$$

Here $\left\langle \hat{\mathbf{F}}(t)\hat{\mathbf{F}}(0) \right\rangle_R$ represents a thermal average of the fluctuation forces in the (stationary) system at a separation $R$ between the relatively moving subsystems. Particularly, the linear-velocity friction force acting on the moving body is $F = -\gamma V$. Equation (12) is a modification of the Kubo formula. As the authors claim, even approximate evaluation of Eq. (12) for the case of van der Waals coupling is difficult. Therefore, they were compelled to introduce many approximations and to consider simplified models. Particularly, in the case $T = 0$ the friction coefficient for an atom moving parallel to the surface is $\gamma \propto \alpha(0)^2 / z_0^{10}$. According to the recent work of Barton [11], who has examined the friction problem using the quantum perturbation theory, the corresponding tangential dissipative force is velocity-proportional only in the fourth order and is unlikely to be of practical interest. In addition, this force is dominated by the term $F \propto V / z_0^8$ and depends sensitively on the atomic line-shapes.

## 2.4 Marvin & Toigo

An important step in the problem of dynamical fluctuation-electromagnetic interaction was done by Marvin and Toigo [12] . They were the first who tried to attack the corresponding relativistic problem. As in the work by Mahanty [14], these authors have used the basic Eq. (11), but the corresponding dispersion equation $D(\omega)$ has been derived when solving the Maxwell equations involving a moving point-dipole source with the appropriate boundary conditions. Unfortunately, the authors have finally restricted their consideration assuming $\beta = V/c << 1$. The obtained expression for the van der Waals atom-surface energy (Eq.(3.7) in Ref.[12]) partly resembles our result [10, 17] when assuming zero temperature of the system and small values of $\beta$. The main difference is that the frequency in the polarizability of the moving particle ( Eq.(3.7) in [12]) is not Doppler-shifted, whereas in the Fresnel amplitudes for reflection of electromagnetic modes the frequency is assumed to be Doppler-shifted. In the nonrelativistic and nonretarded limit the obtained van der Waals potential reduces to (10).

**2.5 Annett & Echenique (1986,1987)**

Annett and Echenique have calculated the real and imaginary parts of the van der Waals potential of a neutral atom moving near a surface using the self-energy formalism [13]. The self-energy is defined by

$$\sum(\mathbf{r},\mathbf{r}',\omega) = i\int \frac{d\omega'}{2\pi} W(\mathbf{r},\mathbf{r}',\omega')G(\mathbf{r},\mathbf{r}',\omega+\omega')\exp(i\delta\omega') \qquad (13)$$

where $G$ is the atomic Green function, $W$ is the screened interaction and $\delta$ is a positive infinitesimal. In terms of this self-energy the van der Waals energy is given by

$$U(z_0) = \iint u_0(\mathbf{r})\sum(\mathbf{r},\mathbf{r}',\omega_0)u_0(\mathbf{r}')d^3r d^3r' \qquad (14)$$

where $u_0$ is the atomic ground-state wave function and summation over spin degeneracy is included. Using Eq. (14), the transition to the dynamical case can be performed in line with that was done by Ferrell and Ritchie [7]. This leads to the expression (in units $e = \hbar = m = 1$)

$$U(z_0) = -\sum_n \frac{f_{n0}}{\omega_{n0}} \int_0^\infty \frac{d\omega}{2\pi} \int d^2q \frac{q^2 \exp(-2qz_0)}{2\pi q} \frac{\text{Im}[g(q,\omega)]}{\omega + \omega_{n0} - \mathbf{V}\cdot\mathbf{q} + q^2/2M - i\cdot 0} \qquad (15)$$

where $f_{n0}$ is the oscillator strength for the atomic dipole transition from state 0 to state $n$, $\omega_{n0} = \omega_n - \omega_0$, $g(q,\omega)$ is the surface response function in the plasmon-pole approximation, $M$ is the particle mass. For low atomic velocities $U(z_0)$ is real and in the simplest case without surface-plasmon dispersion Eq. (15) reduces to (10). This implies that there is no friction in agreement with [15,18]. The friction appears at sufficiently high atomic velocities when the denominator in Eq. (15) becomes zero. This means that the particle energy losses are due to the generation of plasmons in the substrate. The corresponding dissipative force can then be expressed in terms of self-energy $\Sigma$. It is known however, that apart from the one-particle excitations of the surface, there can exist the other, such as for example, the electron-hole pairs where the velocity threshold is absent and which were not taken into account in the theory [13].

**2.6 Other works**

Among the other works we should briefly mention the studies by Levitov [19] and Polevoi [20], and more recently –by Hoye and Brevik [21], Mkrtchian [22], Barton [23], Liebsch [24], Dorofeyev et. al. [25], Scheel and Buhmann [26] devoted to friction forces between the moving bodies. These studies are very different both on the used methods and the results. The last three investigations refer to configuration 1 which is of most interest in the scope of the present work, the other refer to configuration 2. Apart from [21], where the authors have considered a special problem concerning the interaction between the moving oscillators, and [26], where a special case of atomic friction was considered with account of decay rates of atomic states, the main feature in all other works is the lack of friction forces in the nonretarded limit $c \to \infty$. This principally disagrees with subsequent extensive studies performed by several authors (see Sec. 3). Another feature is the lack of totally relativistic final expressions for the friction forces: the main object of calculation is the velocity-linear dissipative force (retarded or nonretarded) and friction coefficient. In addition, we draw attention to the fact that the power of friction force $-\mathbf{F} \cdot \mathbf{V}$ has been frequently (erroneously) identified with the heat losses $dQ/dt$ [19,20,25], with radiation losses [23], or with the Joule dissipation integral [5]. This is incorrect at least in the case of configuration 1, where the correct relationship was later shown to be as follows [17,27]

$$\int \langle \mathbf{j} \cdot \mathbf{E} \rangle d^3 r = dQ/dt + \mathbf{F} \cdot \mathbf{V} \tag{16}$$

Here $\mathbf{F}$ is the fluctuation-electromagnetic force, $\mathbf{j}$ and $\mathbf{E}$ are the current density and electric field in the volume of the particle, $dQ/dt$ is the particle heating rate, all the values being taken in the resting reference frame of the surface. Unfortunately, the existence of Eq. (16) has not received due attention and has led to many contradictions [5, 19, 20, 25].

**3. Modern theoretical studies**

**3.1 Tomassone & Widom (1997)**

Tomassone and Widom [28] have performed detailed nonretarded calculation of the coefficients of friction for charges, dipole molecules and neutral atoms. For the first two types of particles they used the Coulomb Green's functions and the fluctuation-dissipation theorem for the correlator of the fluctuation fields of the surface. The lateral force was then calculated using Eq. (12). However, when performing an analogous calculation for the fluctuating atomic dipole, the authors applied a modification of the nonstationary perturbation theory (without sufficient justification), in which the squared matrix element of the Hamiltonian of the interaction was replaced by the squared matrix element of the operator of the force that acts on the dipole from the fluctuating field of the surface. The final result for the friction force on a neutral atom (in the low-velocity approximation) is given by

$$F = \frac{3\hbar V}{2\pi z_0^5} \int_0^\infty d\omega \alpha''(\omega) \Delta''(\omega) \frac{d}{d\omega} \frac{1}{\exp(\hbar\omega/k_B T) - 1} \qquad (17)$$

where $T$ is the surface temperature, $\alpha''$ is the imaginary part of the atomic polarizability, $\Delta''$ is the imaginary part of the local surface-response function $\Delta(\omega) = (\varepsilon(\omega) - 1)/(\varepsilon(\omega) + 1)$. From (17) it follows that the friction force is zero at $T = 0$. In summary, despite that this work is characterized by the absence of a general theoretical basis which allows to calculate both conservative and dissipative components of fluctuation force within a unified approach, Eq. (17) turns out to be correct, as it has been later proven in our works [17, 27]. Thus, Eq (17) can be considered as an important reference result.

### 3.2 Pendry (1997, 2010)

Pendry has calculated the nonretarded dissipative force in configuration 2, considering the case of two relatively moving perfectly smooth featureless surfaces separated by a finite distance (Fig. 3), and assuming the temperature to be zero [29]. However, without solving the general electrodynamic problem, the author used a heuristic expression for the electrical field in the gap consisting of two parts: (1) the contribution of the fluctuation-electromagnetic field from the immobile plate; and (2) the contribution of the electromagnetic wave reflected from the moving (with velocity $V$) plate, which takes into account the Doppler shift of the frequency of the Fresnel reflection coefficient $R_{1PP}(\omega + k_x V)$ (subscript 1 marks the first of two plates). The corresponding wave has the polarization of P type with the electric vector lying in the same plane that contains the vector normal to the surface and the wave vector. The waves with S-type polarization are not considered owing to nonrelativistic character of the problem and neglecting magnetic properties of media. After the substitution of the amplitude of the resulting field into the tensor of stresses and integration with respect to the projections of the wave vectors $k_x, k_y$ the following expression was obtained for the contact stress applied to the immobile plate from the mobile one (Eq.(18) in [29])

$$F_x/S = \frac{\hbar}{2\pi^3} \int_0^\infty dk_x k_x \int_0^\infty dk_y \exp(-2kl) \int_0^\infty d\omega \left[ R''_{1PP}(\omega + k_x V) - R''_{1PP}(\omega - k_x V) \right] R''_{2PP}(\omega) \qquad (18)$$

where $S$ is the interface area and double primed quantities denote imaginary components Particularly, in the case of linear-velocity expansion Eq. (18) reduces to

$$F_x = \frac{3\hbar SV}{16\pi l^4} \int_0^\infty d\omega R''_{2PP}(\omega) \frac{d}{d\omega} R''_{1PP}(\omega) \tag{19}$$

Note that in the designations which are used in what follows (see also Sec.2.2 and Eq. (17)) we use $\Delta_{1,2}(\omega) \equiv R_{1,2PP}(\omega) = (\varepsilon_{1,2}(\omega)-1)/(\varepsilon_{1,2}(\omega)+1)$, where the indexes 1,2 mark the plates. However, the author [29] did not obtain Eq. (19), since Eq. (18) was not considered as the final one. Instead he made an additional symmetrization in Eq. (18). For this purpose, the order of plates in (18) was formally changed, and the force was written as a half the sum of (18) and the modified expression. As a result, the following expression was obtained (omitting the multiple scattering factor in the denominator which was added later)

$$F_x/S = \frac{\hbar}{2\pi^3} \int_0^\infty dk_x k_x \int_{-\infty}^\infty dk_y \exp(-2kl) \int_0^{k_x V} d\omega \Delta''_1(k_x V - \omega)\Delta''_2(\omega) \tag{20}$$

Formula (20) predicts a precisely zero linear-velocity friction force at $T=0$ in agreement with Eq. (17), referring to configuration 1. The next-order velocity expansion term reads $F_x \propto V^3$. Note that numerical coefficient in (20) is twice as less as that in the recent paper by Pendry [30]. Equation (20) proved to be in agreement with subsequent studies in both configuration 1 (see, for instance, Eq.(4.30) in [27]) and 2 [31-35], but disagree with recent results by Philbin and Leonhardt [36] (see Sec. 3.5).

Despite that Eq. (20) was proven within nonrelativistic quantum formalism, as well [29, 30], the used approach has an obvious lack of generality. In this relation, it is expedient to draw attention to the ambiguities which appear in calculating the attractive force between the plates using the combination of the fields from immobile and mobile plates. Thus, Eq. (15) in [29] for the van der Waals force does not follow from (9),(10) (in [29]) and correct expression includes the product of real and imaginary parts of the corresponding reflection factors with minus sign, i.e.

$$F_z/S = -\frac{\hbar}{(2\pi)^3} \int_{-\infty}^{+\infty} dk_x \int_{-\infty}^{+\infty} dk_y \int_{-\infty}^{+\infty} d\omega k \exp(-2kl)\Delta'_1(\omega + k_x V)\Delta''_2(\omega) \tag{21}$$

Eq. (21) has two obvious drawbacks: 1) incorrect structure of the frequency integrand; 2) minus sign indicating that normal force is (formally) repulsive. Therefore, in the coordinate system used (the second plate is at rest, as shown in Fig.3) the force component $F_x$ acting on plate 2 is obtained with correct (positive) sign, whereas the component $F_z$ is not.

At this point, our general comment referring to this and many other works [6,7,13 14] (see also below) is that an overwhelming majority of researchers have employed the simplest recipe $\omega \to \omega \pm k_x V$ to modify the dielectric permittivity of a moving body when passing from the static situation to the dynamic one for both configurations 1 and 2. Though this recipe makes it possible to obtain correct results in some cases, it does not guarantee that the errors will not appear in other cases.

**3.3 Volokitin and Persson (VP) (1998-2010)**

These authors have considered the dynamical problem in a variety of works [31-35]. So, in [31] they applied a dynamical modification of Lifshitz theory for finding the amplitude of the electric field in the gap between the bodies. Their initial equations contain retardation effects, but since the relative motion of the plates under consideration is slow, some additional simplifying assumptions were made. In these works, the conservative and dissipative forces are treated in a completely identical manner and the tangential (dissipative) force arises as a result of the relative motion of the bodies due to the Doppler shifts in the field amplitudes. The Lifshitz solution of the electrodynamical problem for a resting plate (configuration 2) is modified in the following way: (1) recipe $\omega \to \omega \pm k_x V$ is used in the Fresnel coefficients of a moving plate; (2) the absorption coefficients for P- and S-polarized electromagnetic waves are used as multipliers to the spectral densities of surface-wave modes ($k^2 < \omega^2/c^2$) of fluctuation-electromagnetic field, despite that the used Lifshitz's solution is only valid in the case of thermodynamic equilibrium in the system. As a result, the following expression for the dissipative force in the limit $V/c \ll 1$ is obtained (in our designations for the reflection factors)

$$F_x(l) = -\frac{\hbar S}{16\pi^3} \int_0^\infty d\omega \int_{k<\omega/c} d^2k\, k_x \cdot \left[\frac{(1-|\Delta_{1e}(\omega^+)|^2)(1-|\Delta_{2e}(\omega)|^2)}{|1-\exp(2i|q_0|l)\Delta_{1e}(\omega^+)\Delta_{2e}(\omega)|^2} + (e \leftrightarrow m)\right] \cdot$$

$$\cdot \left[\coth\frac{\hbar\omega}{2k_B T} - \coth\frac{\hbar\omega^+}{2k_B T}\right] -$$

$$-\frac{\hbar S}{4\pi^3} \int_0^\infty d\omega \int_{k>\omega/c} d^2k\, k_x \exp(-2q_0 l)\left[\coth\frac{\hbar\omega}{2k_B T} - \coth\frac{\hbar\omega^+}{2k_B T}\right] \cdot$$

$$\cdot \left[\frac{\mathrm{Im}\,\Delta_{1e}(\omega^+)\,\mathrm{Im}\,\Delta_{2e}(\omega)}{|1-\exp(-2q_0 l)\Delta_{1e}(\omega^+)\Delta_{2e}(\omega)|^2} + (e \leftrightarrow m)\right]$$

(22)

where $\omega^+ = \omega + k_x V$, $q_0 = \sqrt{k^2 - \omega^2/c^2}$, $q_i = \sqrt{k^2 - \varepsilon_i(\omega)\omega^2/c^2}$, $\Delta_{ie} = \frac{q_0\varepsilon_i(\omega) - q_i}{q_0\varepsilon_i(\omega) + q_i}$, $\Delta_{im} = \frac{q_0 - q_i}{q_0 + q_i}$, $i = 1,2$.

The first term in (22) is out of sense in the considered equilibrium situation [37,38]. The reason is that the surface-wave (radiation) part of the fluctuation-electromagnetic field of the plate that is in thermal equilibrium with vacuum background (Lifshitz's situation) has an oscillating (not propagating !) character near the plate irrespective of its velocity.

On the contrary, the second term in (22) corresponding to the near field modes ($k > \omega/c$) provides correct nonrelativistic limit at $c \to \infty$ and takes the form

$$F_x = -\frac{\hbar S}{4\pi^3} \int_0^\infty d\omega \int_{-\infty}^{+\infty} dk_x k_x \int_{-\infty}^{+\infty} dk_y \exp(-2kl) \left[ \coth\frac{\hbar\omega}{2k_B T} - \coth\frac{\hbar\omega^+}{2k_B T} \right]$$
$$\cdot \left[ \frac{\mathrm{Im}\Delta_{1e}(\omega^+)\mathrm{Im}\Delta_{2e}(\omega)}{\left|1-\exp(-2qkl)\Delta_{1e}(\omega^+)\Delta_{2e}(\omega)\right|^2} + (e \leftrightarrow m) \right] \qquad (23)$$

At $T = 0$ formula (23) agrees with (20) and [29,37, 38]. Subsequently it has been reproduced in many works by VP and in their reviewing paper [34], as well, but for a long time VP have not considered the involved dynamical attractive force $F_z$.

Only 10 years after their first works [31] VP have attacked the general relativistic problems in configurations 1 and 2 [35], calculating the basic quantities characterizing relativistic ($V/c \sim 1$) dynamical fluctuation-electromagnetic interaction (tangential and normal forces and radiation heat flux), and assuming the bodies to have different temperatures $T_1$ and $T_2$. Unfortunately, the most part of these results turned out to be in error (see [37,38] for more details). As a characteristic example, we consider below the simplest case of van der Waals attractive force in configuration 1, acting on a resting spherical particle. Using the general result [35], we retrieve the following expression for the resting particle (in our designations)

$$F_z(z_0) = -\frac{\hbar}{2\pi^2}\mathrm{Im}\int_0^\infty d\omega \int_{-\infty}^{+\infty} dk_x \int_{-\infty}^{+\infty} dk_y \exp(-2q_0 z_0) \cdot \left[ \coth\frac{\hbar\omega}{2k_B T_1} + \coth\frac{\hbar\omega}{2k_B T_2} \right] \cdot$$
$$\cdot \left[ \alpha_e(\omega)(k^2 - \omega^2/c^2)\Delta_e(\omega) + \alpha_m(\omega)(k^2 - \omega^2/c^2)\Delta_m(\omega) \right] \qquad (24)$$

where $T_1$ and $T_2$ are the particle and plate temperatures, $\alpha_{e,m}(\omega)$ are the electric (e) and magnetic (m) particle polarizabilities, and amplitudes for Fresnel's refelection coefficients for P- and S-polarized waves are given by (with account of magnetic properties of the plate material)

$$\Delta_e(\omega) = \frac{q_0\varepsilon(\omega)-q}{q_0\varepsilon(\omega)+q}, \quad \Delta_m(\omega) = \frac{q_0\mu(\omega)-q}{q_0\mu(\omega)+q}, \quad q_0 = \sqrt{k^2 - \omega^2/c^2}, \quad q = \sqrt{k^2 - \varepsilon(\omega)\mu(\omega)\omega^2/c^2} \qquad (25)$$

Contrary to (24), correct expression for the Casimir-Polder force has the form [37,38]

$$F_z(z_0) = -\frac{\hbar}{2\pi^2}\int_0^\infty d\omega \int_{-\infty}^{+\infty} dk_x \int_{-\infty}^{+\infty} dk_y \exp(-2q_0 z_0) \cdot$$

$$\left\{ \alpha_e''(\omega)\text{Re}\left(\exp(-2q_0 z_0)[\Delta_e(\omega)(2k^2 - \omega^2/c^2) + \Delta_m(\omega)\omega^2/c^2]\right)\coth\frac{\hbar\omega}{2k_B T_1} + \right.$$

$$+ \alpha_e'(\omega)\text{Im}\left(\exp(-2q_0 z_0)[\Delta_e(\omega)(2k^2 - \omega^2/c^2) + \Delta_m(\omega)\omega^2/c^2]\right)\coth\frac{\hbar\omega}{2k_B T_2} + \quad (26)$$

$$+ \alpha_m''(\omega)\text{Re}\left(\exp(-2q_0 z_0)[\Delta_m(\omega)(2k^2 - \omega^2/c^2) + \Delta_e(\omega)\omega^2/c^2]\right)\coth\frac{\hbar\omega}{2k_B T_1} +$$

$$\left. \alpha_m'(\omega)\text{Im}\left(\exp(-2q_0 z_0)[\Delta_m(\omega)(2k^2 - \omega^2/c^2) + \Delta_e(\omega)\omega^2/c^2]\right)\coth\frac{\hbar\omega}{2k_B T_2} \right\}$$

Comparing (24) and (26) we note the following principal differences:

(1) In Eq. (24) all factors $(2k^2 - \omega^2/c^2)$ are replaced by $(k^2 - \omega^2/c^2)$; (2) Eq. (24) is taken with the erroneous temperature factor $\frac{1}{2}\left(\coth\frac{\hbar\omega}{2k_B T_1} + \coth\frac{\hbar\omega}{2k_B T_2}\right)$, and as a result, at $T_1 \neq T_2$ Eq. (26) can not be written as the imaginary part of analytic function of frequency; (3) As follows from (24), for nonmagnetic particle, $\alpha_m(\omega) = 0$ the attractive Casimir-Polder force does not depend on S-wave contribution of fluctuation-electromagnetic field (the factor $\Delta_m(\omega)$ drops out), and vice versa: for entirely magnetic particle, $\alpha_e(\omega) = 0$ the force does not depend on P-wave contribution (the factor $\Delta_e(\omega)$ drops out). This implies that dynamic generalization of Lifshitz's theory [35] is in principal contradiction to its origin, since Eq. (26) at $T_1 = T_2 = T$ exactly corresponds to the Casimir-Polder force in configuration 1 [37,38], and can be obtained from the Lifshitz formula [2,3] for the attraction force between two plates assuming the limit transition of rarified medium for one plate. Just this limit transition has been used in [35] when deriving (24) from the general relativistic expression for $F_z$ (see Eq. (31) in [35]).

The expression for the radiation heat flux [35] is also in error. So, the corresponding nonrelativistic limit of this expression contains (in the nominator of the integrand) the "ordinary" frequency $\omega$ instead of the Doppler-shifted frequency $\omega^+$. As we have shown in [37,38], this seemingly diminutive error conflicts with the second law of thermodynamics.

To summarize, in contrast to what has been claimed in [31-35] (and here we agree with Leonhardt [41]), the works by VP contain fatal mathematical mistakes and partially contradict each other. This has led to the principally incorrect expression for the tangential force component related to the propagating electromagnetic modes and incorrect expression for the normal force. Only Eq. (23) proves to be correct.

## 3.4 Exact solution to the problem in configuration 1

### *3.4.1 Starting expressions*

Let us consider the case where a small neutral particle moves adiabatically at a constant velocity **V** in vacuum near the smooth surface of a medium (Fig. 4). We assume that the particle is subjected to an electromagnetic field produced by external sources (e.g., laser radiation or quasistatic fields) and to fluctuating fields produced by the heated medium and/or an equilibrium vacuum background (photon gas). The components of the electromagnetic field **E** and **B** satisfy the Maxwell equations and the appropriate boundary conditions. For definiteness, the velocity of the particle is assumed to be parallel to the surface in Fig. 4. However, the expressions presented in this section are valid for any direction of **V**.

In general, the electromagnetic Lorentz force acting on a particle (all the values are given in a laboratory reference frame associated with the surface) is given by

$$\mathbf{F} = \int \rho \mathbf{E}\, d^3 r + \frac{1}{c}\int \mathbf{j} \times \mathbf{B}\, d^3 r \qquad (27)$$

where $\rho$ and **j** are the local charge density and the local current density in the bulk of the particle, respectively, and the integrals are taken over its volume. If all quantities in Eq. (27) undergo statistical fluctuations, then the integrands are treated as statistical averages.

We characterize a small particle by the electric and/or magnetic dipole operators $\mathbf{d}, \mathbf{m}$ respectively, which can be arbitrary functions of time. The dipole approximation is valid if $R/z_0 \ll 1$, where $R$ is the characteristic size (radius) of the particle and $z_0$ is the distance from the surface (Fig. 4). The electric and magnetic polarization vectors produced by the moving particle are

$$\mathbf{P}(\mathbf{r},t) = \mathbf{d}(t)\delta(\mathbf{r} - \mathbf{V}t),\ \mathbf{M}(\mathbf{r},t) = \mathbf{m}(t)\delta(\mathbf{r} - \mathbf{V}t) \qquad (28)$$

With Eqs. (2), the charge and current densities can be written as

$$\rho = -\mathrm{div}\mathbf{P},\ \mathbf{j} = \frac{\partial \mathbf{P}}{\partial t} + c\cdot\mathrm{rot}\mathbf{M} \qquad (29)$$

Using the Maxwell equations $\mathrm{rot}\mathbf{E} = -\frac{1}{c}\frac{\partial \mathbf{B}}{\partial t}$, $\mathrm{div}\mathbf{B} = 0$, and substituting Eqs. (29) into Eq. (27), we perform integration in (27) and transform it to the form

$$\mathbf{F} = \nabla(\mathbf{d}\cdot\mathbf{E} + \mathbf{m}\cdot\mathbf{B}) + \frac{1}{c}\frac{\partial}{\partial t}(\mathbf{d}\times\mathbf{B}) + \frac{1}{c}(\mathbf{V}\nabla)(\mathbf{d}\times\mathbf{B}) = \nabla(\mathbf{d}\cdot\mathbf{E} + \mathbf{m}\cdot\mathbf{B}) + \frac{1}{c}\frac{d}{dt}(\mathbf{d}\times\mathbf{B}) \qquad (30)$$

An analogous expression for the Lorentz force was obtained in [42, 43]. If all the quantities in Eq. (30) undergo fluctuations, then the terms on the right-hand side of Eq. (30) should be subjected to total quantum-statistical averaging. In the case of stationary fluctuations, to which we restrict ourselves in this paper, we can interchange the orders of differentiation with respect to time and statistical averaging in the second term of Eq. (30). As a result, taking into account that $\left\langle \frac{d}{dt}(\mathbf{d}\times\mathbf{B})\right\rangle = \frac{d}{dt}\left\langle \mathbf{d}\times\mathbf{B}\right\rangle = 0$, Eq.(30) takes a more compact form

$$\mathbf{F} = \left\langle \nabla(\mathbf{d}\cdot\mathbf{E}+\mathbf{m}\cdot\mathbf{B})\right\rangle \tag{31}$$

Now, let us calculate the heating (cooling) rate of the particle by the electromagnetic field and/or fluctuating electromagnetic field. For this purpose, using Eqs. (28), (29) and the Maxwell equation $\mathrm{rot}\mathbf{E} = -\frac{1}{c}\frac{\partial\mathbf{B}}{\partial t}$, we transform the integral of dissipated electromagnetic-field energy taken over the volume of the particle in the laboratory reference frame:

$$\int \mathbf{j}\cdot\mathbf{E}\, d^3r = \int \frac{\partial\mathbf{P}}{\partial t}d^3r + c\int \mathbf{E}\cdot \mathrm{rot}\mathbf{M}\, d^3r = (\dot{\mathbf{d}}\cdot\mathbf{E}+\dot{\mathbf{m}}\cdot\mathbf{B})+\mathbf{V}\cdot\nabla(\mathbf{d}\cdot\mathbf{E}+\mathbf{m}\cdot\mathbf{B})-\frac{d}{dt}(\mathbf{m}\cdot\mathbf{B}) =$$
$$= (\dot{\mathbf{d}}\cdot\mathbf{E}+\dot{\mathbf{m}}\cdot\mathbf{B})+\mathbf{V}\cdot\mathbf{F}-\frac{d}{dt}(\mathbf{m}\cdot\mathbf{B})-\frac{\mathbf{V}}{c}\frac{d}{dt}\mathbf{d}\times\mathbf{B} \tag{32}$$

where dots over the dipole moment vectors indicate differentiation with respect to time. For fluctuating moments and electromagnetic fields, the corresponding terms on the right-hand side of Eq. (6) should by statistically averaged, as is done in Eq. (30). In the case of stationary fluctuations, the last two terms are zero and we have

$$\int \left\langle \mathbf{j}\cdot\mathbf{E}\right\rangle d^3r = \left\langle \dot{\mathbf{d}}\cdot\mathbf{E}+\dot{\mathbf{m}}\cdot\mathbf{B}\right\rangle + \mathbf{V}\cdot\left\langle \nabla(\mathbf{d}\cdot\mathbf{E}+\mathbf{m}\cdot\mathbf{B})\right\rangle \tag{33}$$

With Eq. (31), the second term in Eq. (33) is the power produced by the fluctuation tangential force, while the first term is identified with the heating rate of the particle, $dQ/dt$. Therefore,

$$\int \left\langle \mathbf{j}\cdot\mathbf{E}\right\rangle d^3r = \left\langle \dot{\mathbf{d}}\cdot\mathbf{E}+\dot{\mathbf{m}}\cdot\mathbf{B}\right\rangle + \mathbf{V}\cdot\mathbf{F} = \frac{dQ}{dt}+\mathbf{V}\cdot\mathbf{F} \tag{34}$$

Equation (34) is physically clear: the work done by fluctuating electromagnetic field on the moving particle goes into its heating and deceleration. For the first time, this equation was derived in [44] for the case where the magnetic moment of the particle in its rest frame is zero. However, the expression for the heating rate $dQ/dt = \left\langle \dot{\mathbf{d}}\cdot\mathbf{E}+\dot{\mathbf{m}}\cdot\mathbf{B}\right\rangle$, as follows from the above derivation (see also Eq. (31)), holds true irrespective of the value of the magnetic moment in the rest frame of the particle.

Let us derive Eq. (34) using another way, which permits us to show that relativistic transformations for heat and temperature in the Planck relativistic thermodynamics [45] hold true in the case under consideration. Obviously, the integral of Joule heat taken over the volume of the particle in its rest frame is identically equal to the heating rate

$$\frac{dQ'}{dt'} = \int \langle \mathbf{j}' \cdot \mathbf{E}' \rangle d^3 r' \tag{35}$$

Performing standard relativistic transformations of the current density, electric field, and volume in Eq. (35) and taking into account Eq. (33), the right-hand side of Eq. (35) can be recast in the form

$$\int \langle \mathbf{j}' \cdot \mathbf{E}' \rangle d^3 r' = \gamma^2 \left( \int \langle \mathbf{j} \cdot \mathbf{E} \rangle d^3 r - \mathbf{F} \cdot \mathbf{V} \right) \tag{36}$$

where $\gamma = (1 - V^2/c^2)^{-1/2}$ is the Lorentz factor. Following the Planck's relativistic thermodynamics, we have $dQ'/dt = \gamma^2 dQ/dt$. Therefore, combining Eqs. (35) and (36), we obtain Eq. (34) and, taking into account Eqs. (31) and (34), the heating rate of the particle in the laboratory reference frame can be found to be

$$dQ/dt = \langle \dot{\mathbf{d}} \cdot \mathbf{E} + \dot{\mathbf{m}} \cdot \mathbf{B} \rangle \tag{37}$$

Equations (31) and (37) are the most convenient for calculating the conservative and dissipative components of fluctuation-electromagnetic forces and the heating rate of a particle moving near a surface. For this purpose, the terms in the right-hand sides of Eqs. (31) and (37) are represented as sums of pairwise products of spontaneous and induced components

$$F_x = \langle \partial_x (\mathbf{d}^{sp} \mathbf{E}^{in} + \mathbf{d}^{in} \mathbf{E}^{sp} + \mathbf{m}^{sp} \mathbf{B}^{in} + \mathbf{m}^{in} \mathbf{B}^{sp}) \rangle \tag{38}$$

$$F_z = \langle \partial_z (\mathbf{d}^{sp} \mathbf{E}^{in} + \mathbf{d}^{in} \mathbf{E}^{sp} + \mathbf{m}^{sp} \mathbf{B}^{in} + \mathbf{m}^{in} \mathbf{B}^{sp}) \rangle \tag{39}$$

$$\dot{Q} = \langle \dot{\mathbf{d}}^{sp} \mathbf{E}^{in} + \dot{\mathbf{d}}^{in} \mathbf{E}^{sp} + \dot{\mathbf{m}}^{sp} \mathbf{B}^{in} + \dot{\mathbf{m}}^{in} \mathbf{B}^{sp} \rangle \tag{40}$$

The sources of spontaneous fluctuations are: $\mathbf{d}^{sp}$ and $\mathbf{m}^{sp}$---electric and magnetic dipole moments of a moving particle, $\mathbf{E}_s^{sp}$ and $\mathbf{B}_s^{sp}$---spontaneous fluctuating electric and magnetic fields of a surface, $\mathbf{E}_0^{sp}$ and $\mathbf{B}_0^{sp}$---electric and magnetic fields of equilibrium background radiation, filling the space. The induced components of dipole moments and fields are indicated by superscript "$in$". It is worth noting once again, that all the quantities in Eqs. (38)—(40) are given in a laboratory the reference frame associated with the surface.

In the more general case where the fluctuations are not stationary, the expression for the heating rate of the particle, in contrast to Eq. (37), additionally contains the last two terms in the bottom line of Eq. (32) and the expression for the force **F** additionally contains the second term in the right hand side of Eq. (30).

### *3.4.2 General method for calculating fluctuation-electromagnetic forces and rate of particle heating*

Direct statistical averaging in Eqs. (38)—(40) with account of all interactions of the particle interactions with the plate and vacuum background was first performed in our works [27, 44]. For more technical details see also [46, 47]. Using Eqs. (38)—(40), the fluctuation-electromagnetic force and heating rate of the particle are calculated following a scheme which consists of several standard stages.

1. All vector quantities in the right hand sides of Eqs. (38)—(40) except those corresponding to the interaction with vacuum background are represented as Fourier integrals in frequency and in two-dimensional wave vector **k** parallel to the surface of plate (Fig. 4). In the case of background, a three-dimensional integral expansion in **k** is used.

2. A general electrodynamic problem is solved involving calculation of the induced electric and magnetic fields created by the fluctuating dipole moments of the moving particle. According to Eq. (28), in determining the Fourier transforms of polarization $\mathbf{P}_{\omega\mathbf{k}}$ and magnetization $\mathbf{M}_{\omega\mathbf{k}}$ one must obtain the expressions for the spontaneous dipole moments $\mathbf{d}^{sp}(t), \mathbf{m}^{sp}(t)$ in the reference frame $\Sigma$, related with the resting plate (laboratory frame). This is performed using relativistic transformations of these values from the particle rest frame $\Sigma'$ to the laboratory system $\Sigma$:

$$\mathbf{d} = \mathbf{d}' + \frac{1}{c}[\mathbf{V}\,\mathbf{m}'] - \frac{(\gamma-1)}{\gamma}\frac{\mathbf{V}(\mathbf{V}\cdot\mathbf{d}')}{V^2} \tag{41}$$

$$\mathbf{m} = \mathbf{m}' - \frac{1}{c}[\mathbf{V}\,\mathbf{d}'] - \frac{(\gamma-1)}{\gamma}\frac{\mathbf{V}(\mathbf{V}\cdot\mathbf{m}')}{V^2} \tag{42}$$

In finding the Fourier expansions for $\mathbf{d}^{sp}(t), \mathbf{m}^{sp}(t)$ one should first write the corresponding expressions in the system $\Sigma'$. In doing this, the frequency $\omega'$ and time $t'$ are expressed through $\omega$ and $t$ in the reference frame $\Sigma$, while the obtained relations are then substituted into Eqs. (41), (42). Determination of the Fourier transforms of induced electric and magnetic fields generated by moving particle is based on the solution of the system equations for the Fourier transforms of Hertz vectors

$$\left(\Delta + \frac{\omega^2}{c^2}\varepsilon(\omega)\mu(\omega)\right)\Pi^{e}_{\omega\mathbf{k}} = -\frac{4\pi}{\varepsilon(\omega)}\mathbf{P}_{\omega\mathbf{k}} \qquad (43)$$

$$\left(\Delta + \frac{\omega^2}{c^2}\varepsilon(\omega)\mu(\omega)\right)\Pi^{m}_{\omega\mathbf{k}} = -\frac{4\pi}{\mu(\omega)}\mathbf{M}_{\omega\mathbf{k}} \qquad (44)$$

In the case of interaction with vacuum background, Eqs. (43), (44) become algebraic since use is made of three-dimensional expansions in $\mathbf{k}$. Knowing the Fourier transforms of Hertz vectors, the transforms of electric and magnetic field are determined according to the formulas [48]

$$\mathbf{E}_{\omega\mathbf{k}} = \mathrm{graddiv}\,\Pi^{e}_{\omega\mathbf{k}} + \frac{\omega^2}{c^2}\varepsilon(\omega)\mu(\omega)\Pi^{e}_{\omega\mathbf{k}} + \frac{i\omega}{c}\mu(\omega)\mathrm{rot}\,\Pi^{m}_{\omega\mathbf{k}} \qquad (45)$$

$$\mathbf{H}_{\omega\mathbf{k}} = \mathrm{graddiv}\,\Pi^{m}_{\omega\mathbf{k}} + \frac{\omega^2}{c^2}\varepsilon(\omega)\mu(\omega)\Pi^{m}_{\omega\mathbf{k}} - \frac{i\omega}{c}\varepsilon(\omega)\mathrm{rot}\,\Pi^{e}_{\omega\mathbf{k}} \qquad (46)$$

Based on Eqs. (45), (46), the boundary conditions for $\Pi^{e}_{\omega\mathbf{k}}$ and $\Pi^{m}_{\omega\mathbf{k}}$ follow from the continuity conditions for tangential projections of $\mathbf{E}$ and $\mathbf{B}$ at $z = 0$ (see Appendix A).

3. The statistical averages are found corresponding to the interactions between the spontaneous moments of the particle and induced fields of the plate. The involved correlators of dipole moments are calculated using the fluctuation-dissipation relations taken in the particle rest frame $\Sigma'$ [3]

$$\left\langle d_i^{sp'}(\omega)d_k^{sp'}(\omega')\right\rangle = 2\pi\delta_{ik}\delta(\omega+\omega')\hbar\alpha_e''(\omega)\coth\frac{\hbar\omega}{2k_B T_1} \qquad (47)$$

$$\left\langle m_i^{sp'}(\omega)m_k^{sp'}(\omega')\right\rangle = 2\pi\delta_{ik}\delta(\omega+\omega')\hbar\alpha_m''(\omega)\coth\frac{\hbar\omega}{2k_B T_1} \qquad (48)$$

**4.** Calculation of induced moments of the particle in the system $\Sigma$. For this purpose, the corresponding values are expressed in terms of the fluctuating fields through linear integral relations [49] (all the values are taken in the system $\Sigma'$)

$$\mathbf{d}^{ind'}(t') = \int_{-\infty}^{t'}\alpha_e(t'-\tau')\mathbf{E}^{sp'}(\mathbf{r}';\tau')d\tau' \qquad (49)$$

$$\mathbf{m}^{ind'}(t') = \int_{-\infty}^{t'}\alpha_m(t'-\tau')\mathbf{H}^{sp'}(\mathbf{r}';\tau')d\tau' \qquad (50)$$

The induced moments in the system $\Sigma$ are then found using relativistic transforms of electric and magnetic fields in (49), (50) and substituting the obtained dipole moments into Eqs. (41), (42).

5. Calculation of the statistical averages caused by interaction between the fluctuation fields of the plate (vacuum background) and induced dipole moments of the particle is as follows. The correlators of the Fourier components of electric and magnetic field are expressed through the corresponding spectral densities according to the known relations which hold for stationary electromagnetic fluctuations [3]

$$\langle U_i^{sp}{}_{\omega \mathbf{k}} V_j^{sp}{}_{\omega' \mathbf{k}'} \rangle = (2\pi)^4 \delta(\omega+\omega')\delta(\mathbf{k}+\mathbf{k}') \left(U_i^{sp} V_j^{sp}\right)_{\omega \mathbf{k}} \tag{51}$$

where $U_i^{sp}, V_i^{sp} = E_i^{sp}, B_i^{sp}$ (i, j = x, y, z). In their turn, the spectral densities in Eq. (51) are expressed through the spectral densities of the retarded Green function in an isotropic homogeneous and nonmagnetic medium. So, in the case of the plate, the spectral densities are given by [3]

$$\left(E_i^{sp}(z) E_k^{sp}(z')\right)_{\omega \mathbf{k}} = \frac{i}{2} \coth \frac{\hbar \omega}{2 k_B T} \frac{\omega^2}{c^2} \left[D_{ik}(\omega \mathbf{k}, z\, z') - D^*_{ki}(\omega \mathbf{k}, z'z)\right], \tag{52}$$

$$\left(B_i^{sp}(z) B_k^{sp}(z')\right)_{\omega \mathbf{k}} = \frac{i}{2} \coth \frac{\hbar \omega}{2 k_B T} rot_{il} rot'_{km} \begin{bmatrix} D_{lm}(\omega \mathbf{k}, z\, z') - \\ - D^*_{ml}(\omega \mathbf{k}, z'\, z) \end{bmatrix}, \tag{53}$$

$$\left(E_i^{sp}(z) B_k^{sp}(z')\right)_{\omega \mathbf{k}} = \frac{i}{2} \coth \frac{\hbar \omega}{2 k_B T} \frac{i\omega}{c} rot'_{km} \begin{bmatrix} D_{im}(\omega \mathbf{k}, z\, z') - \\ - D^*_{mi}(\omega \mathbf{k}, z'z) \end{bmatrix}, \tag{54}$$

while in the case of vacuum background ($\varepsilon(\omega) = 1 + i \cdot sign(\omega), \mu(\omega) = 1$) Eqs. (52)—(54) reduce to

$$\left(E_i^{sp} E_k^{sp}\right)_{\omega \mathbf{k}} = -\coth \frac{\hbar \omega}{2 k_B T} \frac{\omega^2}{c^2} \operatorname{Im} D_{ik}(\omega, \mathbf{k}) \tag{55}$$

$$\left(B_i^{sp} B_k^{sp}\right)_{\omega \mathbf{k}} = -\coth \frac{\hbar \omega}{2 k_B T} rot_{il} rot'_{km} \operatorname{Im} D_{lm}(\omega, \mathbf{k}) \tag{56}$$

$$\left(E_i^{sp} B_k^{sp}\right)_{\omega \mathbf{k}} = -\coth \frac{\hbar \omega}{2 k_B T} \frac{i\omega}{c} rot'_{km} \operatorname{Im} D_{lm}(\omega, \mathbf{k}) \tag{57}$$

Note that Eqs. (52)—(54) are written in terms of the "surface" representation for the components of Green's function (vector $\mathbf{k}$ is two-dimensional), while Eqs. (55)—(57) are written in terms of the "volume" representation for the components of Green's function (vector $\mathbf{k}$ is three-dimensional). A complete set of Green's functions, including their surface and vacuum parts at thermal equilibrium between the plate and vacuum background is given in Appendix B. The expressions for the Fourier transforms of Hertz vectors and the corresponding components of electric and magnetic fields (induced in the plate and in the background) are given in [43, 44].

6. Using the aforementioned results and Eqs. (38)—(40), the final expressions for the fluctuation-electromagnetic forces and heating rate are found.

*3.4.3 Nonrelativistic case*

It is expedient to consider a simpler nonrelativistic problem for illustrating the general method in more detail. Neglecting for simplicity the magnetic terms in Eqs. (38)—(40), our starting expressions take the form

$$F_x = \langle \partial_x (\mathbf{d}^{sp} \mathbf{E}^{in}) \rangle + \langle \partial_x (\mathbf{d}^{in} \mathbf{E}^{sp}) \rangle \tag{58}$$

$$F_z = \langle \partial_z (\mathbf{d}^{sp} \mathbf{E}^{in}) \rangle + \langle \partial_z (\mathbf{d}^{in} \mathbf{E}^{sp}) \rangle \tag{59}$$

$$\dot{Q} = \langle \dot{\mathbf{d}}^{sp} \mathbf{E}^{in} \rangle + \langle \dot{\mathbf{d}}^{in} \mathbf{E}^{sp} \rangle \tag{60}$$

The calculation of the first terms in (58)--(60) describing the contributions from $\mathbf{d}^{sp}$ is reduced to the solution of Poisson's equation with fluctuating point dipole source

$$\Delta \phi = 4\pi \, div [\delta(x - Vt)\delta(y)\delta(z - z_0) \mathbf{d}^{sp}(t)] \tag{61}$$

Equation (30) has to be solved taking into account the continuity conditions for the potential $\phi$ and normal component of the electric induction at $z = 0$. The calculated Fourier transform for the induced part of $\phi$ is then given by

$$\Phi^{in}_{\omega \mathbf{k}}(z) = \frac{2\pi}{k} \Delta(\omega)[ik_x d^{sp}_x (\omega - k_x V) + ik_y d^{sp}_y (\omega - k_x V) + k d^{sp}_z (\omega - k_x V)] \exp(-k(z + z_0)) \tag{62}$$

where $\Delta(\omega) = (\varepsilon(\omega) - 1)/(\varepsilon(\omega) + 1)$ and $d^{sp}_{x,y,z}(\omega - k_x V)$ are the projections of the Fourier transform of $\mathbf{d}^{sp}(t)$. Using Eq. (61), the induced electric field is given by $\mathbf{E}^{in} = -\nabla \phi^{in}$.

Further, the Fourier expansions of $\mathbf{E}^{in}$ and $\mathbf{d}^{sp}$ are substituted into the first terms of Eqs. (58)—(60). The statistical averaging in the appeared correlators of dipole moments is carried out using Eq. (47), but prior to this we should differentiate the corresponding expressions with respect to coordinates $x, z$ (in (58), (59)) and with respect to time $t$ (in (60)), and substitute the particle coordinates ($x = Vt, y = 0, z = z_0$) in the involved integrands. Finally, after integration over the frequencies and components of wave vector we obtain (omitting the lower index at $z_0$ for simplicity)

$$F_x^{(1)}(T_1) = \frac{\hbar}{\pi^2} \int_0^\infty d\omega \int_{-\infty}^{+\infty} dk_x \int_{-\infty}^{+\infty} dk_y \, k k_x \exp(-2kz) \Delta''(\omega) \alpha_e''(\omega^+) \coth\left(\frac{\hbar \omega^+}{2k_B T_1}\right) \tag{63}$$

$$F_z^{(1)}(T_1) = -\frac{\hbar}{\pi^2} \int_0^\infty d\omega \int_{-\infty}^{+\infty} dk_x \int_{-\infty}^{+\infty} dk_y \, k^2 \exp(-2kz) \Delta'(\omega) \alpha_e''(\omega^+) \coth\left(\frac{\hbar \omega^+}{2k_B T_1}\right) \tag{64}$$

$$\dot{Q}^{(1)}(T_1) = -\frac{\hbar}{\pi^2} \int_0^\infty d\omega \int_{-\infty}^{+\infty} dk_x \int_{-\infty}^{+\infty} dk_y \, k \exp(-2kz) \Delta''(\omega) \alpha_e''(\omega^+) \omega^+ \coth\left(\frac{\hbar \omega^+}{2k_B T_1}\right) \tag{65}$$

where $\omega^+ = \omega + k_x V$, and one-primed and double-primed functions denote the corresponding real and imaginary components. As one can see, Eq. (65) follows from (63) if we replace $k_x$ by $-\omega^+$. This is caused by differentiation with respect to $x$ in (58), and with respect to $t$ in (60). The Doppler-shifted frequency arises as a consequence of the change of the induced electric field in the rest frame of the particle. This obvious fact is not taken into account in [35], where the formula for the heating rate contains the nonshifted frequency $\omega$ (see also Sect. 3.3).

To find the contribution from the induced moments of the particle (the second terms in (58)—(60)), Eq. (49) is used. Then we have

$$\mathbf{d}^{ind}(t) = \int_{-\infty}^{t} d\tau\, \alpha_e(t-\tau) \mathbf{E}^{sp}(V\tau, 0, z_0, \tau) = $$
$$= \frac{1}{(2\pi)^3} \iiint d\omega\, d^2k\, \alpha_e(\omega - k_x V) \mathbf{E}^{sp}_{\omega\mathbf{k}}(z_0) \exp(-i(\omega - k_x V)t) \quad (66)$$

In writing Eq. (66), the field of the surface is taken at the point of the particle localization. When substituting Eq. (66) and the Fourier expansion of $\mathbf{E}^{sp}$ into the second terms of (58)—(60), the arising correlators are simplified using Eqs. (51), (52).

As it follows from Maxwell's equations, the electric field created by neutral particle with dipole moment $\mathbf{d}(t) = \mathbf{d}(\omega)\exp(-i\omega t)$ located at the point $\mathbf{r}'$ obeys the equation [3]

$$\left(rotrot - \varepsilon(\omega)\omega^2/c^2\right)\mathbf{E}(\omega,\mathbf{r}) = 4\pi(\omega^2/c^2)\mathbf{d}(\omega)\delta(\mathbf{r}-\mathbf{r}') \quad (67)$$

Comparing Eq. (1) for retarded Green's function with Eq. (67) we see that the spectral Green's function $D_{lm}(\omega,\mathbf{r},\mathbf{r}')$ at fixed $m$ and $\mathbf{r}'$ is equal to the electric field produced by the point-like dipole at the point $\mathbf{r}'$

$$d_l(\omega) = -\frac{\hbar c^2}{\omega^2}\delta_{lm}\,(l, m = x, y, z) \quad (68)$$

In the nonrelativistic approximation, the solution of Eq. (67) reduces to the solution of Eq. (61). Using (62) and (68) yields the Green functions and taking into account (52) we obtain the corresponding spectral densities

$$\left(E_x^{sp}(z_0) E_x^{sp}(z_0)\right)_{\omega\mathbf{k}} = \hbar \coth\frac{\hbar\omega}{2k_B T} \frac{2\pi}{k} k_x^2 \exp(-2kz_0)\,\mathrm{Im}\,\Delta(\omega) \quad (69)$$

$$\left(E_y^{sp}(z_0) E_y^{sp}(z_0)\right)_{\omega\mathbf{k}} = \hbar \coth\frac{\hbar\omega}{2k_B T} \frac{2\pi}{k} k_y^2 \exp(-2kz_0)\,\mathrm{Im}\,\Delta(\omega) \quad (70)$$

$$\left(E_z^{sp}(z_0)E_z^{sp}(z_0)\right)_{\omega\mathbf{k}} = \hbar\coth\frac{\hbar\omega}{2k_BT}\frac{2\pi}{k}k^2\exp(-2kz_0)\,\mathrm{Im}\,\Delta(\omega) \tag{71}$$

As well, according to (51) and (69)—(71) we have

$$\left\langle \mathbf{E}^{sp}_{\omega\mathbf{k}}(z_0)\mathbf{E}^{sp}_{\omega'\mathbf{k}'}(z_0)\right\rangle = 2(2\pi)^4 k\exp(-2kz_0)\hbar\coth(\hbar\omega/2k_BT)\,\mathrm{Im}\,\Delta(\omega)\delta(\omega+\omega')\delta(\mathbf{k}+\mathbf{k}') \tag{72}$$

Making use of Eq. (72) and performing trivial integrations over $\omega'$ and $\mathbf{k}'$ in the corresponding expressions for the components of fluctuation-electromagnetic forces and heating rate we obtain

$$F_x^{(2)}(T_2) = -\frac{\hbar}{\pi^2}\int_0^\infty d\omega\int_{-\infty}^{+\infty}dk_x\int_{-\infty}^{+\infty}dk_y\,kk_x\exp(-2kz)\Delta''(\omega)\alpha_e''(\omega^+)\coth\left(\frac{\hbar\omega}{2k_BT_2}\right) \tag{73}$$

$$F_z^{(2)}(T_2) = -\frac{\hbar}{\pi^2}\int_0^\infty d\omega\int_{-\infty}^{+\infty}dk_x\int_{-\infty}^{+\infty}dk_y\,k^2\exp(-2kz)\Delta'(\omega)\alpha_e''(\omega^+)\coth\left(\frac{\hbar\omega}{2k_BT_2}\right) \tag{74}$$

$$\dot{Q}^{(2)}(T_2) = \frac{\hbar}{\pi^2}\int_0^\infty d\omega\int_{-\infty}^{+\infty}dk_x\int_{-\infty}^{+\infty}dk_y\,k\exp(-2kz)\Delta''(\omega)\alpha_e''(\omega^+)\omega^+\coth\left(\frac{\hbar\omega}{2k_BT_2}\right) \tag{75}$$

Finally, collecting Eqs. (63)—(65) and (73)—(75) yields

$$F_x = -\frac{\hbar}{\pi^2}\int_0^\infty d\omega\int_{-\infty}^{+\infty}dk_x\int_{-\infty}^{+\infty}dk_y\,kk_x\exp(-2kz)\Delta''(\omega)\alpha_e''(\omega^+)\left[\coth\left(\frac{\hbar\omega}{2k_BT_2}\right)-\coth\left(\frac{\hbar\omega^+}{2k_BT_1}\right)\right] \tag{76}$$

$$F_z = -\frac{\hbar}{\pi^2}\int_0^\infty d\omega\int_{-\infty}^{+\infty}dk_x\int_{-\infty}^{+\infty}dk_y\,k^2\exp(-2kz)\left[\begin{array}{l}\Delta''(\omega)\alpha_e'(\omega^+)\coth\left(\dfrac{\hbar\omega}{2k_BT_2}\right)+\\ +\Delta'(\omega)\alpha_e''(\omega^+)\coth\left(\dfrac{\hbar\omega^+}{2k_BT_1}\right)\end{array}\right] \tag{77}$$

$$\dot{Q} = \frac{\hbar}{\pi^2}\int_0^\infty d\omega\int_{-\infty}^{+\infty}dk_x\int_{-\infty}^{+\infty}dk_y\,k\exp(-2kz)\Delta''(\omega)\alpha_e''(\omega^+)\omega^+\left[\coth\left(\frac{\hbar\omega}{2k_BT_2}\right)-\coth\left(\frac{\hbar\omega^+}{2k_BT_1}\right)\right] \tag{78}$$

We have first obtained Eqs. (33)--(35) in [27] in the different (equivalent) form. The present form seems to be more natural in view of the nonrelativistic limit $c\to\infty$ for the corresponding relativistic expressions of $F_x, F_z, dQ/dt$ (Sect. 3.4.4). The form of Eqs. (33)--(35) where the integration over $k_x, k_y$ is performed using symmetric limits $(-\infty,+\infty)$ is more convenient when passing from configuration 1 to configuration 2 (Sect. 3.6).

A brief analysis of Eqs. (33)--(35) reveals several features. The presence of Doppler-shifted frequencies $\omega^+$ in arguments of some functions, as it may seem, enables to guess the right final expression, at least for the dynamical van der Waals force, if use is made of the Doppler-shifted frequencies in the corresponding variables. However, as it has been demonstrated in Section 2, correct result (Eq.(8)) has not previously been found in this way. As far as concerned the tangential force $F_x$, a possibility to get correct result using this approach seems to be even more problematic. First, since $F_x = 0$ at $V = 0$, we have no starting formula. Second, it is very difficult to guess the Doppler-shifted frequency before the temperature factor in Eq. (35) without finding the general solution to the electrodynamic problem (even having correct static expression for $dQ/dt$ at $V = 0$). In this relation, we must stress once again, that the basic method for solving dynamical problems in both configurations 1 and 2 is the fluctuation electrodynamics in the form given by Lifshitz and Pitaevskii 50 years ago [2,3] despite it has been mostly applied to static problems.

Second, it is worthwhile to consider the temperature dependence of the van der Waals force when passing from the equilibrium state $T_1 = T_2$ to the nonequilibrium one, $T_1 \neq T_2$. The presence of two independent sources of spontaneous fluctuations in (28) and subsequent averaging using the fluctuation-dissipative relations (31) and (32) manifests that the dependence $F_z$ on $T_1$ and $T_2$ may have only the form of Eq. (34), where $\text{Im}\Delta(\omega)$ is multiplied by $\coth(\hbar\omega/2k_B T_2)$ and $\text{Im}\alpha(\omega^+)$---by $\coth(\hbar\omega^+/2k_B T_2)$, correspondingly. This is also true (with certain complications) in the case of relativistic problem, as well. Therefore, by no means correct expression for $F_z$ can not include common temperature factor with or without Doppler-shifted frequency (such as [35]), if the problem would been correctly solved.

Now consider several limiting cases which follow from (76)—(78). So, in linear velocity expansion at $T_1 = T_2 = T$ Eq. (76) reduces to (17)---the result by Tomassone and Widom [28]. In the cold limit $T_1 = T_2 = T \to 0$, taking into account that $\coth(\hbar\omega/2k_B T) \to sign(\omega)$, we obtain [27]

$$F_x = \frac{4\hbar}{\pi^2} \int_0^\infty dk_x k_x \int_0^\infty dk_y k \exp(-2kz) \int_0^{k_x V} d\omega \alpha''(\omega - k_x V)\Delta''(\omega) =$$
$$= -\frac{4\hbar V}{\pi^2 z^5} \int_0^\infty du (u^3 K_0(2u) + 0.5u^2 K_1(2u)) \int_0^u dp\, \alpha''(\omega_0(u-p))\Delta''(\omega_0 p),\ \omega_0 = V/z$$
(79)

This is the so-called "quantum friction force" [29]. The nonlinear velocity dependence is determined by the certain form of the functions $\alpha''(\omega)$ and $\Delta''(\omega)$. In particular, since the outer integrand function takes maximum values close to $u \sim 1$, the force $F_x$ takes the maximum value if the characteristic absorption frequencies of the particle and surface are close one another and to $\omega_0 = V/z$. At $z = 1\,nm$, for example, the particle velocity will be $10^8\,cm/s$ (i.e. of order Bohr's velocity).

Both Eq. (17) and Eq. (76) describes dissipative tangential force. However, as we have concluded in [27] and quite recently [50], at $T_1 \neq T_2$ the tangential force can be accelerating. This is a principally new result. Fig. 5 illustrates the situation in the case of a $MgO$ nanoparticle (with radius of $1\,nm$) moving above a $SiC$ substrate [50]. We see that the acceleration effect is observed in the certain intervals of velocities ($0.06 \div 0.2 V_B$ at $z = 10\,nm$) and distances ($5 \div 20\,nm$ at $V = 0.2 V_B$). The corresponding formulas are given in Appendix C.

At $V = 0$ formula (77) predicts the nonretarded van der Waals (Casimir-Polder) attraction force with thermal contribution. So, at $T_1 = T_2 = T$ and $V = 0$ Eq. (77) can be recast in the well-known form

$$F_z = -\frac{3}{2}\frac{k_B T}{z^4}\sum_{n=0}^{\infty}(1 - \delta_{n0}/2)\alpha_e(i\xi_n)\Delta(i\xi_n),\ \xi_n = (2\pi k_B T/\hbar)n \qquad (80)$$

In the case $V \neq 0$, $T_1 = T_2 = 0$ Eq. (77) reduces to (8) and (9) in Ref. [9] if we take into account that $F_z = -\partial U/\partial z$, and a single oscillator atomic model is used.

In its turn, Eq. (78) at $V = 0$ reduces to the known expression for the near-field particle-surface radiative heat flux [51,52].

### 3.4.4 Relativistic case

Relativistic calculation of fluctuation-electromagnetic interaction in configuration 1 (Fig. 4) has been carried out in our works [17, 44, 46, 47]. In comparison with the nonrelativistic problem statement, the number of independent sources of spontaneous fluctuation increases now. These are $\mathbf{d}^{sp}$ and $\mathbf{m}^{sp}$ ---the electric and magnetic dipole moments of the moving particle, $\mathbf{E}_s^{sp}$ and $\mathbf{B}_s^{sp}$---spontaneous fluctuating electric and magnetic fields of the surface, and $\mathbf{E}_0^{sp}$, $\mathbf{B}_0^{sp}$---background thermal electromagnetic fields (electric and magnetic). The method of the calculation is in line with Sect. 3.4.2 though technically it becomes more tedious. Due to the lack of correlations between the equilibrium background radiation and fluctuation field of the surface, the involved contributions are found separately from one another. The starting expressions are (38)—(40) including the analogous terms from the vacuum field contributions. In the latter case the vacuum-field contribution to the force $F_z$ is obviously zero. Finally we obtain the following expressions

$$F_x = -\frac{\hbar\gamma}{2\pi^2}\int_0^\infty d\omega \int_{-\infty}^{+\infty} dk_x \int_{-\infty}^{+\infty} dk_y k_x \left[\alpha_e''(\gamma\omega^+)\text{Im}\left(\frac{\exp(-2q_0 z)}{q_0}R_e(\omega,\mathbf{k})\right) + (e\leftrightarrow m)\right] \cdot$$
$$\cdot\left[\coth\left(\frac{\hbar\omega}{2k_B T_2}\right) - \coth\left(\frac{\gamma\hbar\omega^+}{2k_B T_1}\right)\right] - \qquad (81)$$
$$-\frac{\hbar\gamma}{\pi c^4}\int_0^\infty d\omega\,\omega^4 \int_{-1}^{1} dx\, x(1+\beta x)^2 [\alpha_e''(\gamma\omega_1) + \alpha_m''(\gamma\omega_1)] \cdot \left[\coth\left(\frac{\hbar\omega}{2k_B T_2}\right) - \coth\left(\frac{\gamma\hbar\omega_1}{2k_B T_1}\right)\right]$$

$$F_z = -\frac{\hbar\gamma}{2\pi^2}\int_0^\infty d\omega \int_{-\infty}^{+\infty} dk_x \int_{-\infty}^{+\infty} dk_y \left\{\begin{array}{l}\alpha_e''(\gamma\omega^+)\text{Re}[\exp(-2q_0 z)R_e(\omega,\mathbf{k})]\coth\left(\frac{\gamma\hbar\omega^+}{2k_B T_1}\right) + \\ + \alpha_e'(\gamma\omega^+)\text{Im}[\exp(-2q_0 z)R_e(\omega,\mathbf{k})]\coth\left(\frac{\hbar\omega}{2k_B T_2}\right) + (e\leftrightarrow m)\end{array}\right\} \qquad (82)$$

$$\dot{Q} = \frac{\hbar\gamma}{2\pi^2}\int_0^\infty d\omega \int_{-\infty}^{+\infty} dk_x \int_{-\infty}^{+\infty} dk_y\,\omega^+ \cdot \left[\alpha_e''(\gamma\omega^+)\text{Im}\left(\frac{\exp(-2q_0 z)}{q_0}R_e(\omega,\mathbf{k})\right) + (e\leftrightarrow m)\right]$$
$$\cdot\left[\coth\left(\frac{\hbar\omega}{2k_B T_2}\right) - \coth\left(\frac{\gamma\hbar\omega^+}{2k_B T_1}\right)\right] + \qquad (83)$$
$$+\frac{\hbar\gamma}{\pi c^3}\int_0^\infty d\omega\,\omega^4 \int_{-1}^{1} dx(1+\beta x)^3[\alpha_e''(\gamma\omega_1)+\alpha_m''(\gamma\omega_1)]\cdot\left[\coth\left(\frac{\hbar\omega}{2k_B T_2}\right) - \coth\left(\frac{\gamma\hbar\omega_1}{2k_B T_1}\right)\right]$$

where $\omega^+ = \omega + k_x V$, $\omega_1 = \omega(1+\beta x)$, $\beta = V/c$, $\gamma = (1-\beta^2)^{-1/2}$

$$R_e(\omega,\mathbf{k}) = \Delta_e(\omega)\left[2(k^2 - k_x^2\beta^2)(1-\omega^2/k^2c^2) + (\omega^+)^2/c^2\right] +$$
$$+ \Delta_m(\omega)\left[2k_y^2\beta^2(1-\omega^2/k^2c^2) + (\omega^+)^2/c^2\right] \qquad (84)$$

$$R_m(\omega,\mathbf{k}) = \Delta_m(\omega)\left[2(k^2 - k_x^2\beta^2)(1-\omega^2/k^2c^2) + (\omega^+)^2/c^2\right] +$$
$$+ \Delta_e(\omega)\left[2k_y^2\beta^2(1-\omega^2/k^2c^2) + (\omega^+)^2/c^2\right] \qquad (85)$$

$$\Delta_e(\omega) = \frac{q_0\varepsilon(\omega)-q}{q_0\varepsilon(\omega)+q},\quad \Delta_m(\omega) = \frac{q_0\mu(\omega)-q}{q_0\mu(\omega)+q},\quad q = \left(k^2 - (\omega^2/c^2)\varepsilon(\omega)\mu(\omega)\right)^{1/2},$$
$$q_0 = (k^2 - \omega^2/c^2)^{1/2},\quad k^2 = k_x^2 + k_y^2 \qquad (86)$$

In the above expressions $\varepsilon(\omega)$ and $\mu(\omega)$ are the frequency –dependent dielectric permittivity and magnetic permeability of the plate material, $\alpha_e(\omega)$ and $\alpha_m(\omega)$ are the frequency – dependent dipole electric and magnetic polarizabilities of the particle. One primed and double primed quantities represent the corresponding real and imaginary parts, $T_1$ is the particle temperature, $T_2$ is the plate and vacuum-field temperature (both the vacuum background and the plate are in thermal equilibrium). The global system of magnetodielectric bodies is out of thermal equilibrium, but in a stationary regime. It is worth noting that Eqs. (81)—(83) are written in the resting system of the surface (and background). The second terms in Eqs. (81), (83) describes the contributions from the equilibrium photon gas (vacuum electromagnetic field): tangential force and rate of heat exchange.

In the case $V \ll c$ the expression for the tangential friction force applied to a small particle (beyond the limits of geometrical optics approximation) moving through an equilibrium photon gas of temperature $T$ was first obtained by Mkrtchian et. al. [53]. In this case ($T_1 = T, T_2 = T$) the second line of Eq. (81) reduces to

$$F_x = -\frac{\hbar^2 V}{3\pi c^5 k_B T} \int_0^\infty d\omega \omega^5 \frac{\alpha_e''(\omega)}{\sinh^2(\hbar\omega/2k_B T)} \qquad (87)$$

The general solution to the relativistic problem of photonic drug was later obtained in our works [53]. Eqs.(81)—(83) provide exact solution to the relativistic problem in configuration 1 in the dipole approximation of fluctuation-electromagnetic theory, which is valid at any velocities and temperatures $T_1, T_2$, assuming the plate and vacuum background to be in thermal equilibrium. To date, this is a unique configuration for which such a general relativistic solution was obtained, being self-consistent and agreeing with any limiting cases. So, in the limit $c \to \infty$ Eqs. (81)—(83) reduce to Eqs. (76)—(78). At $V = 0, T_1 = T_2 = T$ Eq. (82) describes the Casimir-Polder force with account of thermal contribution [46]

$$F_z = -2k_B T \sum_{n=0}^\infty a_n \int_0^\infty dk k [R_e(i\xi_n, k)\alpha_e(i\xi_n) + R_m(i\xi_n, k)\alpha_m(i\xi_n)] \exp\left(-2\sqrt{k^2 + \xi_n^2/c^2}\,z\right) \qquad (88)$$

where $a_n = (1 - \delta_{0n}/2)$ and $\xi_n = 2\pi k_B T n/\hbar$.

### 3.5 Philbin and Leonhardt (2009, 2010)

The situation with dynamical generalization of Lifshitz' theory in configuration 2 has become even more complex after the recent works by Philbin and Leonhardt (PL) [36,41], where it was claimed that the quantum frictional (dissipative) force is precisely zero at $T = 0$, a result in contradiction to many previous works. As well, PL manifest the absence of quantum friction in configuration 1. We agree with the refuting argumentation [54] that PL had omitted to consider the effects of the Doppler shift on the reflection coefficient of the moving surface [36,41]. Owing to this shift, the reflection coefficient of a medium that is moving in the chosen reference frame has an altered analytic structure in the complex frequency plane giving rise to the nonzero dissipative force at $T = 0$. This result is valid for any real medium with nonzero dissipation. We also agree with the refuting argumentation in [37,55], that there is no relativistic invariance for the evanescent modes of the fluctuating electromagnetic field which is viewed in [36,41] as the ground reason for the zero frictional force .

In line with the Lifshitz theory [2,3], PL determine retarded Green function for the system of two plates in relative motion (with different temperatures $T_1$ and $T_2$), and calculate the mean values of the components of stress tensor and Pointing vector to obtain the forces $F_x, F_z$ and the rate of radiative heat flux. Their final expressions resemble to some extent the corresponding results in [35], but differ in some principal points. So, the hyperbolic cotangent of the moving plate in [36] has the modulus of Doppler-shifted frequency for both normal and lateral force components. Owing to this, the friction force is zero at $T_1 = T_2 = 0$ in any order of the velocity expansion. Commenting this result, PL relate it with the Lorentz invariance of the zero-frequency modes of electromagnetic field. However, for any medium with material properties (such as the plates in configuration 2), the zero modes of electromagnetic field are not invariant. Each of the plates can be related to its own inertial reference frame and there is no physical reason for the tangential force to be zero. More formally, the tangential force is obtained when considering nonrelativistic limit of this force in configuration 2, which is given by (see Sect. 3.6 and [37])

$$F_x = -\frac{\hbar S}{4\pi^3}\int_0^\infty d\omega \int_{-\infty}^{+\infty} dk_x k_x \int_{-\infty}^{+\infty} dk_y \exp(-2kl)\frac{\Delta_1''(\omega^+)\Delta_2''(\omega)}{\left|1-\exp(-2kl)\Delta_1(\omega^+)\Delta_2(\omega)\right|^2} \cdot$$
$$\cdot\left[\coth\frac{\hbar\omega}{2k_B T_2}-\coth\frac{\hbar\omega^+}{2k_B T_1}\right] \quad (89)$$

To date, Eq. (89) is one of the reference results (as well as $F_z$ and $\dot{Q}$ at $c \to \infty$, see in what follows) in the scope of configuration 2, and probably it is the unique result where there is an overall concordance between the results of different authors [29,31--35,37].

Making use the limit transition $T_1 \to 0$ and $T_2 \to 0$ in (89),

$$\lim_{T1\to 0}\coth\frac{\hbar\omega^+}{2k_B T_1} = sign(\omega+k_x V), \lim_{T2\to 0}\coth\frac{\hbar\omega}{2k_B T_2} = sign\omega, \quad (90)$$

we obtain by no means the nonzero result

$$F_x(l) = -\frac{\hbar S}{2\pi^3}\int_0^{+\infty} dk_x k_x \int_{-\infty}^{+\infty} dk_y \int_0^{k_x V} d\omega \exp(-2kl)\frac{\Delta_1''(k_x V-\omega)\Delta_2''(\omega)}{\left|1-\exp(-2kl)\Delta_1(\omega-k_x V)\Delta_2(\omega)\right|^2} \quad (91)$$

in complete agreement with (20) and [29]. The opposite sign in (91) (as compared to [29,30]) is due to the fact that we calculate the force applied to the moving plate 1, whereas in [29,30] the force is calculated through the component of Maxwell's stress tensor on the surface of resting plate 2, i.e. the force is applied to plate 2.

If in the second hyperbolic cotangent (46) we replace $\omega^+$ by $|\omega^+|$, the result becomes principally different, and we obtain $F_x = 0$. Generally speaking, there is no any reason for such a replacement, and the frequency argument of the hyperbolic cotangent must be both positive and negative. One may be convinced in this, for example, when calculating the mean value of the squared dipole moment for the resting dipole moment. Thus, using the fluctuation-dissipation relation (31) yields

$$\langle (\mathbf{d}^{sp})^2 \rangle = 3\hbar \int_{-\infty}^{+\infty} \frac{d\omega}{2\pi} \alpha''(\omega) \coth \frac{\hbar\omega}{2k_B T_1} \tag{92}$$

Bearing in mind parity of the integrand in (92) we obtain

$$\langle (\mathbf{d}^{sp})^2 \rangle = \frac{3\hbar}{\pi} \int_0^{+\infty} d\omega\, \alpha''(\omega) \coth \frac{\hbar\omega}{2k_B T_1} \tag{93}$$

At the same time, making use substitution $\coth \frac{\hbar\omega}{2k_B T_1} \to \coth \frac{\hbar|\omega|}{2k_B T_1}$ in (92) we obtain the physically incorrect zero result. In our opinion, PL made this error in the process of transformation of triple integral over the wave vector and frequency.

### 3.6 The correspondence rules between configurations 1,2 and dynamical generalization of Lifshitz's theory in configuration 2

The limiting transition from configuration 2 to configuration 1 using the limit of rarified medium for one of the plates was proposed by Lifshitz in 1955 [2]. Subsequently this approach was used by many other authors [35, 39, 40, 56].

To perform such a transition, the dielectric and magnetic permittivity of the plate chosen are represented in the form $\varepsilon_1(\omega) = 1 + 4\pi n_1 \alpha_e(\omega)$, $\mu_1(\omega) = 1 + 4\pi n_1 \alpha_m(\omega)$ assuming the condition $n_1 \alpha_{e,m}(\omega) \ll 1$ to be fulfilled, and the expression for the attractive force (energy) is then found in the linear-order expansion over this small parameter. As a result, we obtain the following expressions

$$F_z^{(1)}(z_0) = -\frac{1}{n_1 S} \frac{dF_z^{(2)}(l)}{dl}\bigg|_{l=z_0} \tag{94}$$

$$\Delta_{1e}(\omega) \to \frac{\pi n_1}{q_0^2} \left[ \alpha_e(\omega)(2k^2 - \omega^2/c^2) + \alpha_m(\omega)\omega^2/c^2 \right] \tag{95}$$

$$\Delta_{1m}(\omega) \to \frac{\pi n_1}{q_0^2}\left[\alpha_m(\omega)(2k^2-\omega^2/c^2)+\alpha_e(\omega)\omega^2/c^2\right] \qquad (96)$$

where $F_z^{(2)}(l)$ is the attraction force between the two plates, $l$ is the corresponding gap width, $F_z^{(1)}(z_0)$ is the attraction force acting on the particle a distance $z_0$ apart from the plate, and $\alpha_{e,m}$ are the particle electric and magnetic polarizabilities. In the nonrelativistic approximation $c \to \infty$, at $T_1 \neq T_2$ and $V \neq 0$ the corresponding transition rules take the form

$$F_{x,z}^{(1)}(z_0) = -\frac{1}{n_1 S}\frac{dF_{x,z}^{(2)}(l)}{dl}\bigg|l=z_0, \quad \dot{Q}^{(1)}(z_0) = -\frac{1}{n_1 S}\frac{d\dot{Q}^{(2)}(l)}{dl}\bigg|l=z_0 \qquad (97)$$

$$\Delta_{1e}(\omega) \to 2\pi n_1 \alpha_e(\omega), \quad \Delta_{1m}(\omega) \to 2\pi n_1 \alpha_m(\omega) \qquad (98)$$

In the absence of a well recognized selfconsistent solution to the dynamical problem in configuration 2, we have found a surprising possibility to use the exact solution obtained for the configuration 1 when obtaining the corresponding solution for configuration 2. This was done using the correspondence rule (CR) between both dynamical configurations, that was first proposed in our works [17, 37, 38].

The main idea of CR assumes that for each configuration all the values which characterize the fluctuation-electromagnetic interaction ($F_x, F_z, \dot{Q}$) are obtained from the solution of general electrodynamic problem. Owing to the linearity of the limiting transition $1 \to 2$ (97) the sequence of transformations

$$F_z^{(0)} \to F_z \to F_x \to \dot{Q},$$

where $F_z^{(0)}$ is the force of static ($V=0$) van der Waals attraction at thermal equilibrium ($T_1 = T_2 = T$) will necessarily have analogous form in both configurations.

It is convenient to rewrite Eq. (77) for the van der Waals attractive force in configuration 1 (at $V=0, T_1 = T_2 = T$) in the form (omitting the subscript "e" in the polarizability coefficients, for simplicity)

$$F_z = -\frac{\hbar}{\pi^2}\int_0^\infty d\omega \int_{-\infty}^{+\infty} dk_x \int_{-\infty}^{+\infty} dk_y\, k^2 \exp(-2kz)\left[\Delta''(\omega)\alpha'(\omega^+)\coth\left(\frac{\hbar\omega}{2k_B T}\right) + \Delta'(\omega)\alpha''(\omega^+)\coth\left(\frac{\hbar\omega}{2k_B T}\right)\right] \qquad (99)$$

Comparing Eq. (99) with (76)—(78) we see that the latter equations are obtained from (99) via the sequence of transformations

$$\left.\begin{array}{l} F_z^{(0)} \to F_z, \\ \Delta''(\omega)\coth(\hbar\omega/2k_B T) \to \Delta''(\omega)\coth(\hbar\omega/2k_B T_2), \\ \alpha''(\omega)\coth(\hbar\omega/2k_B T) \to \alpha''(\omega^+)\coth(\hbar\omega^+/2k_B T_1), \\ \alpha'(\omega), \alpha''(\omega) \to \alpha'(\omega^+), \alpha''(\omega^+) \end{array}\right\} \quad (100)$$

$$2.\left.\begin{array}{l} F_z \to F_x, \\ d^2k\,k \to d^2k\,k_x \\ \Delta''(\omega) \to \Delta''(\omega), \Delta'(\omega) \to \Delta''(\omega) \\ \alpha'(\omega^+) \to \alpha''(\omega^+), \alpha''(\omega^+) \to -\alpha''(\omega^+) \end{array}\right\} \quad (101)$$

$$3.\left.\begin{array}{l} F_x \to \dot{Q} \\ d^2k\,k_x \to -d^2k\,\omega^+ \end{array}\right\} \quad (102)$$

where $\omega^+ = \omega + k_x V$. On the other hand, according to CR, the analogous sequence of transformations that are completely identical to (100)—(102) should hold in the case of configuration 2, when coming from the static equilibrium state to the dynamical (nonequilibrium) case. In doing this we must substitute $\Delta_1(\omega) = (\varepsilon_1(\omega)-1)/(\varepsilon_1(\omega)+1)$ for $\alpha(\omega)$ (the dielectric function of moving plate 1) and $\Delta_2(\omega) = (\varepsilon_2(\omega)-1)/(\varepsilon_2(\omega)+1)$ for $\Delta(\omega)$ (the dielectric function of resting plate 2). Taking this into account, we obtain the sequence of transformations

$$1.\left.\begin{array}{l} F_z^{(0)}(l) \to F_z(l), \\ \Delta_2''(\omega)\coth(\hbar\omega/2k_B T) \to \Delta_2''(\omega)\coth(\hbar\omega/2k_B T_2), \\ \Delta_1''(\omega)\coth(\hbar\omega/2k_B T) \to \Delta_1''(\omega^+)\coth(\hbar\omega^+/2k_B T_1), \\ \Delta_1'(\omega), \Delta_1''(\omega) \to \Delta_1'(\omega^+), \Delta_1''(\omega^+) \end{array}\right\} \quad (103)$$

$$2.\left.\begin{array}{l} F_z(l) \to F_x(l), \\ d^2k\,k \to d^2k\,k_x \\ \Delta_2''(\omega) \to \Delta_2''(\omega), \Delta_2'(\omega) \to \Delta_2''(\omega) \\ \Delta_1'(\omega^+) \to \Delta_1''(\omega^+), \Delta_1''(\omega^+) \to -\Delta_1''(\omega^+) \end{array}\right\} \quad (104)$$

$$3.\left.\begin{array}{l} F_x(l) \to \dot{Q}(l) \\ d^2k\,k_x \to -d^2k\,\omega^+ \end{array}\right\} \quad (105)$$

The starting expression for the van der Waals attraction force in configuration 2 at $V=0, T_1=T_2=T$ can be cast in the form [17, 37, 38]

$$F_z^{(0)}(l) = -\frac{\hbar S}{4\pi^3}\int_0^\infty d\omega \int_{-\infty}^{+\infty} dk_x \int_{-\infty}^{+\infty} dk_y\, k\, \frac{\exp(-2kl)}{|1-\exp(-2kl)\Delta_1(\omega)\Delta_2(\omega)|^2} \cdot \\ \cdot [\Delta_1''(\omega)\Delta_2'(\omega)\coth(\hbar\omega/2k_B T) + \Delta_1'(\omega)\Delta_2''(\omega)\coth(\hbar\omega/2k_B T)] \quad (106)$$

Making use the sequence of transformations (103)—(105) yields [17,37,38]

$$F_z^{(2)}(l) = -\frac{\hbar S}{4\pi^3}\int_0^\infty d\omega \int_{-\infty}^{+\infty} dk_x \int_{-\infty}^{+\infty} dk_y k \frac{\exp(-2kl)}{\left|1-\exp(-2kl)\Delta_1(\omega^+)\Delta_2(\omega)\right|^2} \cdot$$
$$\cdot \left[\Delta_1''(\omega^+)\Delta_2'(\omega)\coth(\hbar\omega^+/2k_BT_1) + \Delta_1'(\omega^+)\Delta_2''(\omega)\coth(\hbar\omega/2k_BT_2)\right] \quad (107)$$

$$F_x^{(2)}(l) = -\frac{\hbar S}{4\pi^3}\int_0^\infty d\omega \int_{-\infty}^{+\infty} dk_x \int_{-\infty}^{+\infty} dk_y k_x \frac{\exp(-2kl)}{\left|1-\exp(-2kl)\Delta_1(\omega^+)\Delta_2(\omega)\right|^2}\Delta_1''(\omega^+)\Delta_2''(\omega) \cdot$$
$$\cdot \left[\coth(\hbar\omega/2k_BT_2) - \coth(\hbar\omega^+/2k_BT_1)\right] \quad (108)$$

$$\dot{Q}^{(2)}(l) = \frac{\hbar S}{4\pi^3}\int_0^\infty d\omega \int_{-\infty}^{+\infty} dk_x \int_{-\infty}^{+\infty} dk_y \omega^+ \frac{\exp(-2kl)}{\left|1-\exp(-2kl)\Delta_1(\omega^+)\Delta_2(\omega)\right|^2}\Delta_1''(\omega^+)\Delta_2''(\omega) \cdot$$
$$\cdot \left[\coth(\hbar\omega/2k_BT_2) - \coth(\hbar\omega^+/2k_BT_1)\right] \quad (109)$$

If we want to account for magnetic properties of the plates we should add the analogous terms with $\tilde{\Delta}_i(\omega) = (\mu_i(\omega)-1)/(\mu_i(\omega)+1), i=1,2$ into the right-hand sides of (107)—(109). It is not difficult to verify that substituting (107)—(109) in (97), (98) immediately results in Eqs. (76)—(78).

These results can be thought of as nonequilibrium (dynamical and thermal) generalization of Lifshitz's theory in configuration 2 within the scope of nonrelativistic and nonretarded approximation. Formulae (107)—(109) are well justified in the case $V \neq 0$ and $T_1 \neq T_2$.

For two resting plates out of thermal equilibrium (between one another and with surrounding vacuum background, $V = 0, T_1 \neq T_2 \neq T_3$) the problem was recently examined by Antezza et. al. [57], who obtained the expressions for $F_z$ and $\dot{Q}$. It is precisely this work where the authors reached a deeper understanding of the principal role that is played by the background radiation in configuration 2.

As well, at $V \ll c$ with account of retardation but under equilibrium $T_1 = T_2 = T_3$, the problem was examined using the correspondence rule [17,37,38], where we have obtained the expressions for $F_x, F_z, \dot{Q}$. To date, however, solving the general relativistic problem in configuration 2 at any velocities and out of equilibrium ($T_1 \neq T_2 \neq T_3$) remains a challenging puzzle for future investigators.

## 4. A brief review of experimental results

### 4.1 Normal and tangential dynamical fluctuation-electromagnetic forces

The combination of new theoretical elaborations and experimental capabilities at submicronre scale has led to a growing number of experiments that probe the geometry and material dependence of fluctuation –electromagnetic interactions in static regime (see [4,58--61] for a review). To date, however, information on dynamical van der Waals and Casimir-Lifshitz forces remains extremely scarce. There is no experimental information concerning dynamical corrections to the conservative Casimir-Lifshitz forces, while a few attempts in measuring the corresponding dissipative (frictional) components have been reported in [62—66] using dynamical regime of atomic force microscope.

In particular, the damped motion of oscillating tips of atomic force niscroscope (AFM) was studied, but several attempts to interpret these results using the currently used theory of fluctuation-electromagnetic interaction turned out to be unsuccessful [25, 27, 31, 34, 63]. This is partly due to the fact that the viscous friction (proportional to velocity) can be caused by various factors (apart from the damping forces at closer repulsive contacts which are not related to the problem of dynamical fluctuation-electromagnetic forces). A fundamental problem here is that in the noncontact dynamic vacuum mode of an AFM with compensated contact potential difference, the conservative interaction between a tip and a sample is determined by the van der Waals force; therefore, it is reasonable to expect that vacuum friction will have the same origin. However, theoretical estimates showed that the measured damping forces exceed the calculated ones by two to three orders of magnitude for silicon–mica contacts [34, 67] and by 5 to 11 orders of magnitude for metal-metal contacts [27, 34, 63, 68 ].

Experiments also give different dependences of damping forces on the distance from the surface, namely, $z^{-3}$ in [64] and from $z^{-1.1}$ to $z^{-1.5}$ in [65]; in the last case a strong influence of temperature and the type of contacting materials is observed. One should note that in [62--64] the damped oscillations of AFM tip occurred along the normal to the surface, whereas in [65]--- along the surface and the tip radius was greater by a factor of 30 to 50 in the latter case (1 μm). The strongly different power-law dependence of the damping force observed in [65] is likely due to the electrostatic interaction between charged spots rather than to the van der Waals (dissipative) interaction, which is characterized by the $z^{-3}$ dependence [64]. The interpretation of the results of [65] is also hampered by the fact that the attractive force between the tip and the surface was not measured in this case. In [64], on the contrary, there is no doubt that the attractive force is of van der Waals type and a problem is only the small difference between theoretical and experimental values of damping forces [67].

As we have reported [46], the damping forces measured with an AFM in [64] and exhibiting a distance dependence $z^{-3}$ can be explained using the small-velocity limit of Eq. (81) if an AFM tip and the surface have the same absorption band at a frequency about $10^9$ Hz. Such frequencies are typical of rotational excitations of molecular complexes and phonon excitations. Moreover, the inverse decay time of oscillators in experiments with a quartz microbalance is of the same order of magnitude [69--71]. Other explanations of the experimental data from [64, 65] were also proposed [34]; however, the accuracy of dissipative-force measurements is still insufficiently high for choosing the adequate theoretical model. At present, we even cannot state with a confidence that the experimental data [65–65] are related to vacuum friction caused by fluctuation-electromagnetic mechanism rather than to forces of other origin. In the recent paper [66], the authors have measured the friction coefficient of an AFM tip above a Nb film in the transition temperature range from normal to superconducting state (9.2 *K*). In particular, the variations of the friction coefficient with the tip-surface distance (in the pendulum geometry) and the bias voltage were studied. It was found that the friction coefficient decreases by a factor of three when the sample enters the superconducting state. This behavior was explained by relative contribution from two mechanisms of noncontact friction: fluctuation-electromagnetic (electronic) and phononic. Both mechanisms are responsible for the friction in normal state of Nb film, while only the last one in the superconducting state. As well, the authors have observed different distance dependence of the friction coefficient: about $z^{-1}$ and $z^{-4}$ in the two aforementioned cases. Since this interpretation is based on theoretical models [34] which do not provide quantitative agreement between the calculated (fluctuation –electromagnetic) and measured values of the friction coefficient, we cannot assert with a confidence that conflict between the theory and experiment is eliminated. Further studies of damping forces are required differing in terms of the type of contact, velocity, temperature, and the geometrical and mechanical characteristics of tips. From the point of view of possible experimental measurements of dynamical Casimir-Lifshitz forces it is likely more preferable to use nanomechanical resonators or the scattering of fast neutral atoms (clusters) from the surface. The limiting factor in the former case is a rather small velocity of oscillation, while in the latter one---obtaining a high-velocity beam of uncharged molecules (clusters).

### 4.2 Dynamical Casimir effect

Forty years ago, it was suggested [72] that a mirror undergoing relativistic motion could convert virtual photons into directly observable real photons. This phenomenon was later termed the dynamical Casimir effect [73, 74]. Quite recently, the dynamical Casimir effect was observed [75] in a superconducting circuit consisting of a coplanar transmission line with a tunable electrical length. An open transmission line was terminated by a SQUID (superconducting quantum interference device). Since the rate of change of the electrical length can be close to the speed of light by modulating the inductance of the SQUID at high frequencies ($>10$ gigahertz), the electrical field cannot adiabatically adapt to these changes and can be nonadiabatically excited out of the vacuum. However, as we can see, in this experiment we are not dealing with a really moving mirror. For some other alternative proposals see the references cited in [75].

## 5. Conclusions

To conclude, we have shown that an elementary transition from static to dynamic configurations 1,2 in the general problem of fluctuation-electromagnetic interaction that is performed by formal transition to the Doppler-shifted frequency in the dielectric and thermodynamic characteristics of moving bodies (small particle and plate) may lead to incorrect results even in rather simple situations, though some results may prove to be correct. However, a physically clear understanding of these results (as the limiting cases of more general ones) is impossible without a consecutive solution of the corresponding dynamical problems on the base of fluctuation electrodynamics. In more complex situations, the attempts to solve the problem avoiding the straightforward solution turned out to be inconsistent.

To date, the exact solution of the general relativistic problem at any $\beta = V/c$ and $T_1 \neq T_2 = T_3$ has been obtained only for configuration 1 (Eqs. (39)—(41)). For dynamical out of equilibrium configuration 2 the problem has been only solved in the nonrelativistic case (Eqs. (56)—(58)), and within the restricted relativistic statement with account of retardation $\beta << 1, T_1 = T_2 = T_3$, as well [17,37,38]. A solution to the general relativistic problem in configuration 2 (at $\beta \to 1, T_1 \neq T_2 \neq T_3$) still remains an open challenging question.

Experiments relevant to dynamical fluctuation-electromagnetic forces are scarce and insufficiently convincing, despite several attempts which were done to date. More definite situation concerns the dynamical Casimir effect.

## APPENDIX A

Using the continuity conditions for the tangential components of electric and magnetic fields (46),(47) at $z = 0$ results in the continuity conditions for the products of Hertz vectors:

$$\varepsilon(\omega)\mu(\omega)\Pi_i^{(e,m)} \quad (i = x, y), \tag{A1}$$

$$i k_x \Pi^e_x + i k_y \Pi^e_y + \frac{\partial \Pi^e_z}{\partial z}, \tag{A2}$$

$$i k_x \Pi^m_x + i k_y \Pi^m_y + \frac{\partial \Pi^m_z}{\partial z}, \tag{A3}$$

$$\varepsilon(\omega)\frac{\partial \Pi_i^e}{\partial z} \quad (i = x, y), \quad \varepsilon(\omega)\Pi^e_z, \tag{A4}$$

$$\mu(\omega)\frac{\partial \Pi_i^m}{\partial z} \quad (i = x, y) , \quad \mu(\omega)\Pi^m_z \tag{A5}$$

**APPENDIX B**

$$D_{xx}(\omega\mathbf{k}; z\, z') = \left(-\frac{\hbar c^2}{\omega^2}\right)\frac{2\pi}{q_0}\exp(-q_0|z-z'|)\cdot\left(\frac{\omega^2}{c^2}-k_x^2\right)+$$
$$+\left(-\frac{\hbar c^2}{\omega^2}\right)\frac{2\pi}{q_0}\exp(-q_0(z+z'))\left[k_x^2\left(1-\frac{\omega^2}{k^2c^2}\right)\Delta_e(\omega)+\frac{\omega^2}{c^2}\frac{k_y^2}{k^2}\Delta_m(\omega)\right] \tag{B.1}$$

$$D_{yy}(\omega\mathbf{k}; z\, z') = \left(-\frac{\hbar c^2}{\omega^2}\right)\frac{2\pi}{q_0}\exp(-q_0|z-z'|)\cdot\left(\frac{\omega^2}{c^2}-k_y^2\right)+$$
$$+\left(-\frac{\hbar c^2}{\omega^2}\right)\frac{2\pi}{q_0}\exp(-q_0(z+z'))\left[k_y^2\left(1-\frac{\omega^2}{k^2c^2}\right)\Delta_e(\omega)+\frac{\omega^2}{c^2}\frac{k_x^2}{k^2}\Delta_m(\omega)\right] \tag{B.2}$$

$$D_{zz}(\omega\mathbf{k}; z\, z') = \left(-\frac{\hbar c^2}{\omega^2}\right)\frac{2\pi}{q_0}\exp(-q_0|z-z'|)k^2 + \left(-\frac{\hbar c^2}{\omega^2}\right)\frac{2\pi}{q_0}\exp(-q_0(z+z'))k^2\Delta_e(\omega) \tag{B.3}$$

$$D_{xy}(\omega\mathbf{k}; z\, z') = D_{yx}(\omega\mathbf{k}; z, z') = \left(-\frac{\hbar c^2}{\omega^2}\right)\frac{2\pi}{q_0}\exp(-q_0|z-z'|)(-1)k_x k_y +$$
$$+\left(-\frac{\hbar c^2}{\omega^2}\right)\frac{2\pi}{q_0}\exp(-q_0(z+z'))\cdot k_x k_y\left[\left(1-\frac{\omega^2}{k^2c^2}\right)\Delta_e(\omega)-\frac{\omega^2}{k^2c^2}\Delta_m(\omega)\right] \tag{B.4}$$

$$D_{xz}(\omega\mathbf{k}; z\, z') = \left(-\frac{\hbar c^2}{\omega^2}\right)\frac{2\pi}{q_0}\exp(-q_0|z-z'|)\cdot(-q_0)ik_x\,sign(z-z') +$$
$$+\left(-\frac{\hbar c^2}{\omega^2}\right)\frac{2\pi}{q_0}\exp(-q_0(z+z'))\cdot(-q_0)ik_x\,\Delta_e(\omega) \tag{B.5}$$

$$D_{zx}(\omega\mathbf{k}; z\, z') = \left(-\frac{\hbar c^2}{\omega^2}\right)\frac{2\pi}{q_0}\exp(-q_0|z-z'|)\cdot(-q_0)ik_x\,sign(z-z') +$$
$$+\left(-\frac{\hbar c^2}{\omega^2}\right)\frac{2\pi}{q_0}\exp(-q_0(z+z'))\cdot ik_x q_0\,\Delta_e(\omega) \tag{B.6}$$

$$D_{zy}(\omega\mathbf{k}; z\, z') = \left(-\frac{\hbar c^2}{\omega^2}\right)\frac{2\pi}{q_0}\exp(-q_0|z-z'|)\cdot(-q_0)ik_y\,sign(z-z') +$$
$$+\left(-\frac{\hbar c^2}{\omega^2}\right)\frac{2\pi}{q_0}\exp(-q_0(z+z'))\cdot ik_y q_0\,\Delta_e(\omega) \tag{B.7}$$

$$D_{yz}(\omega\mathbf{k}; z\, z') = \left(-\frac{\hbar c^2}{\omega^2}\right)\frac{2\pi}{q_0}\exp(-q_0|z-z'|)\cdot(-q_0)ik_y\,sign(z-z') +$$
$$+\left(-\frac{\hbar c^2}{\omega^2}\right)\frac{2\pi}{q_0}\exp(-q_0(z+z'))\cdot(-q_0)ik_y\,\Delta_e(\omega) \tag{B.8}$$

The first terms in (B.1)—(B.8) correspond to the components of Green's functions for free vacuum space in the "surface" representation. The second terms are caused by the contribution $D^{(S)}_{ik}(\omega\mathbf{k}, z, z')$ from the plate.

## APPENDIX C

To analyze the possibility of accelerative force which follows from Eq.(76), let us consider the nondissipative approximations for the particle and surface dielectric properties:

$$\alpha(\omega) = \frac{\alpha(0)\omega_0^2}{\omega_0^2 - \omega^2 - i \cdot 0 \cdot \text{sign}(\omega)} \quad (C.1)$$

$$\varepsilon(\omega) = 1 - \frac{\omega_P^2}{\omega(\omega + i \cdot 0)} \quad (C.2)$$

From (C.1),(C.2) it follows

$$\alpha''(\omega) = \frac{\pi \alpha(0)\omega_0}{2}\left[\delta(\omega - \omega_0) - \delta(\omega + \omega_0)\right] \quad (C.3)$$

$$\Delta''(\omega) = \frac{\pi \omega_s}{2}\left[\delta(\omega - \omega_s) - \delta(\omega + \omega_s)\right] \quad (C.4)$$

Substituting (C.3),(C.4) into Eq.(76) and performing integration over $\omega$ and $k_y$ yields

$$F_x = -\frac{\hbar \alpha(0)\omega_0 \omega_s}{64 z_0^4}\left\{\frac{1}{(\omega_0 - \omega_s)} x_1^4 [K_0(x_1) + K_2(x_1)] \cdot \left[\coth\left(\frac{\hbar \omega_s}{2k_B T_2}\right) - \coth\left(\frac{\hbar \omega_0}{2k_B T_1}\right)\right] + \frac{1}{(\omega_0 + \omega_s)} x_2^4 [K_0(x_2) + K_2(x_2)] \cdot \left[\coth\left(\frac{\hbar \omega_s}{2k_B T_2}\right) + \coth\left(\frac{\hbar \omega_0}{2k_B T_1}\right)\right]\right\}, \quad (C5)$$

where $x_1 = 2|\omega_0 - \omega_s|z_0/V$, $x_2 = 2(\omega_0 + \omega_s)z_0/V$, $\omega_0$ and $\omega_s$ are characteristic absorption frequencies of the particle and surface material.

## Figure captions

Fig. 1
Model of interacting media and coordinate system.

Fig. 2
Dynamical configuration 1.

Fig. 3
Dynamical configuration 2.

Fig. 4
Dynamical configuration 1. General case.

Fig. 5. The velocity and temperature dependence of tangential forces on a MgO particle ($R = 1nm$) above a SiC substrate. The solid, dotted, dashed and dashed –dotted lines correspond

to the substrate temperatures of 1500, 1200, 900 and 600K. A particle has zero temperature, the distance to the surface equals $10 nm$. The values of velocity are given in Bohr units. (From [50]).

Fig. 6. The distance and temperature dependence of tangential force on a MgO particle ($R = 1nm$) moving with velocity $V = 0.1V_B$ above a SiC substrate. The solid, dotted, dashed and dashed –dotted lines correspond to the substrate temperatures of 1500, 1200, 900 and 600K (from [50]).

Fig. 7. The distance and velocity dependence of tangential force on a MgO particle ($R = 1nm$) above a SiC substrate with temperature 1500K. The solid, dotted, dashed and dashed –dotted lines correspond to the particle velocities of 0.02, 0.04, 0.08 and $0.16V_B$ (from [50]).

FIGURE 1

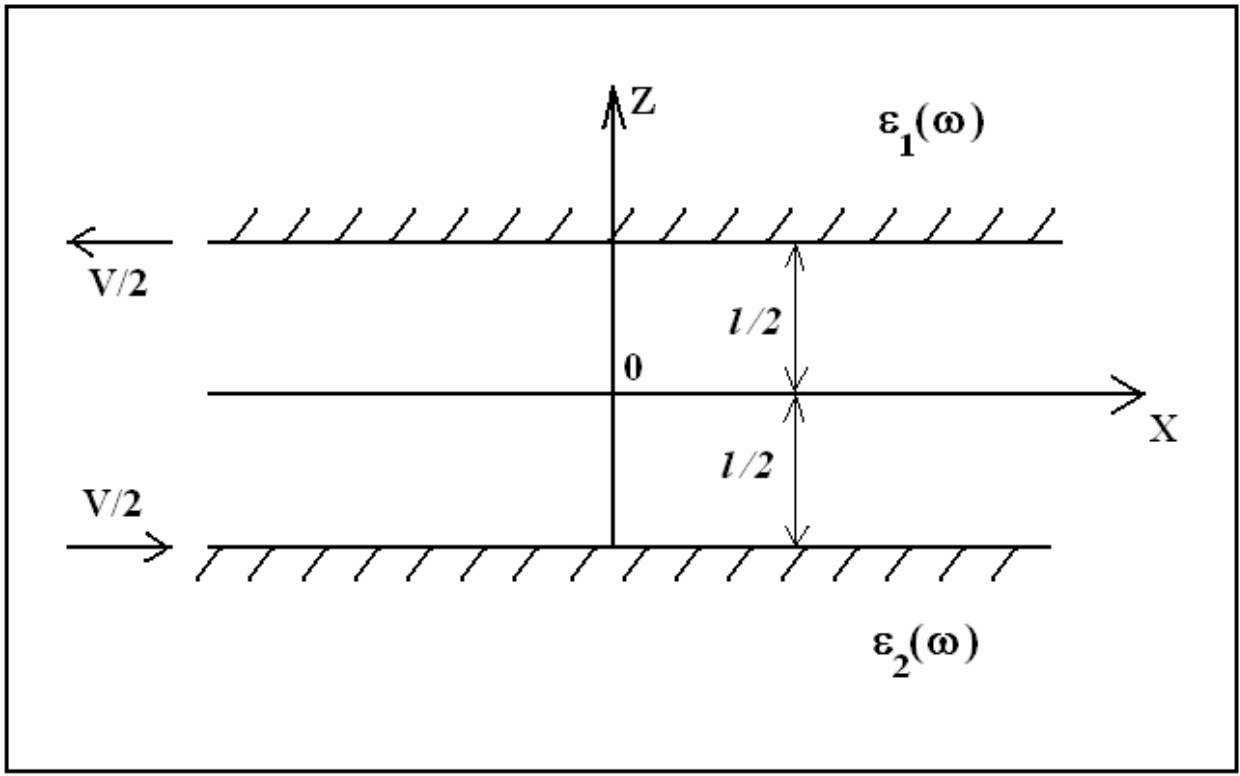

FIGURE 2

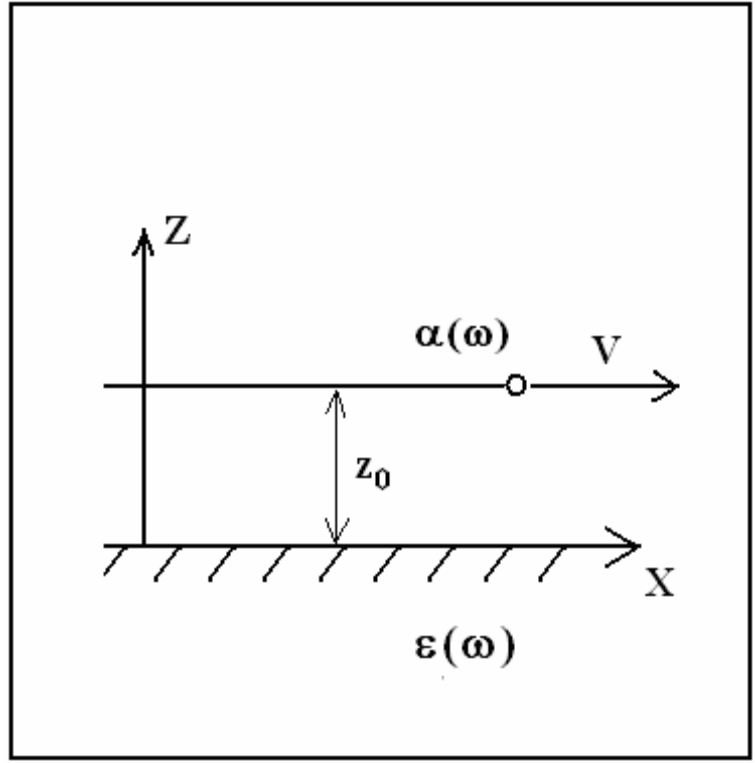

FIGURE 3

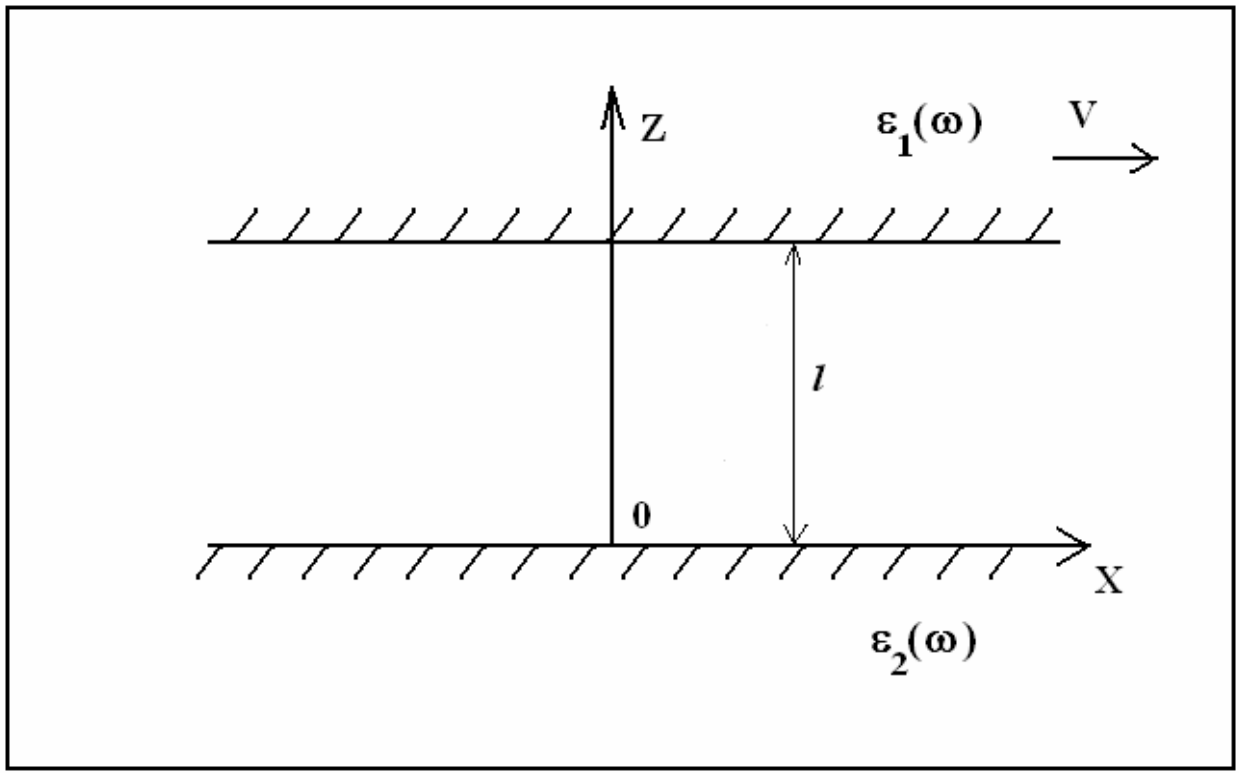

FIGURE 4

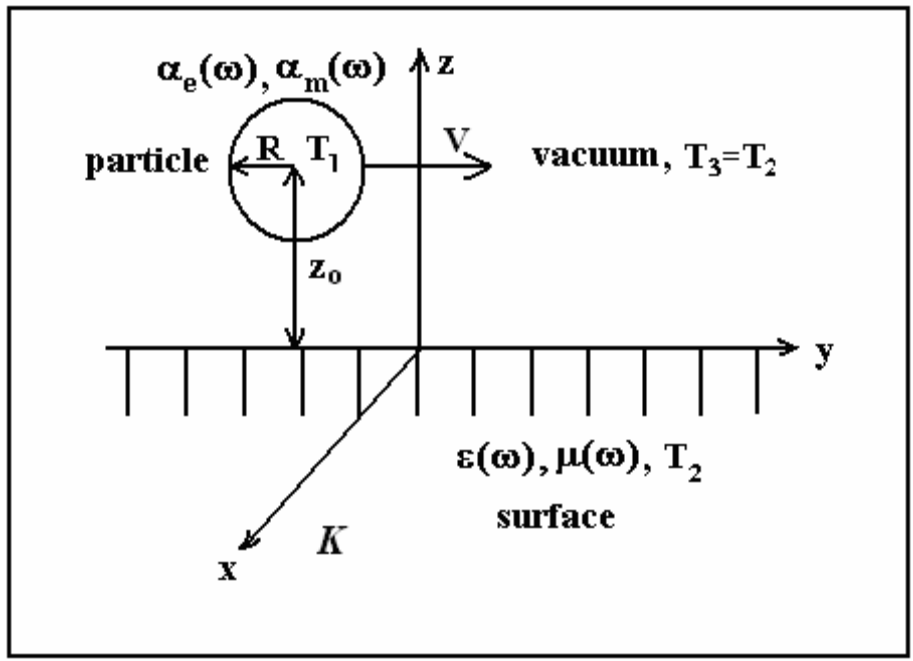

FIGURE 5

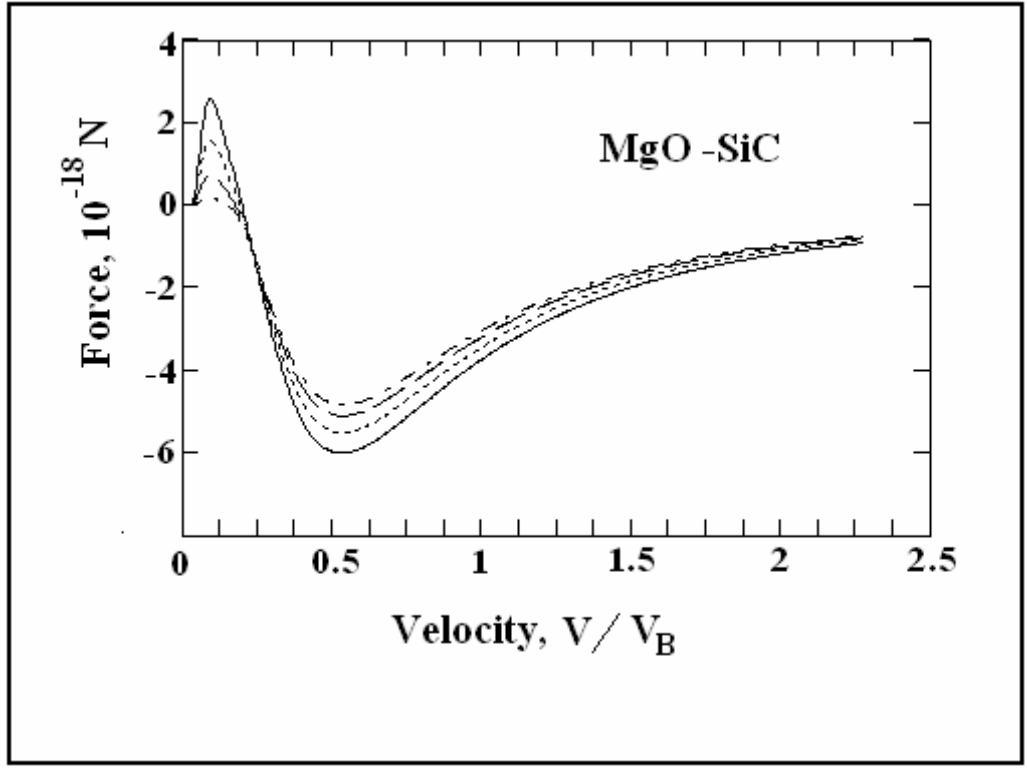

FIGURE 6

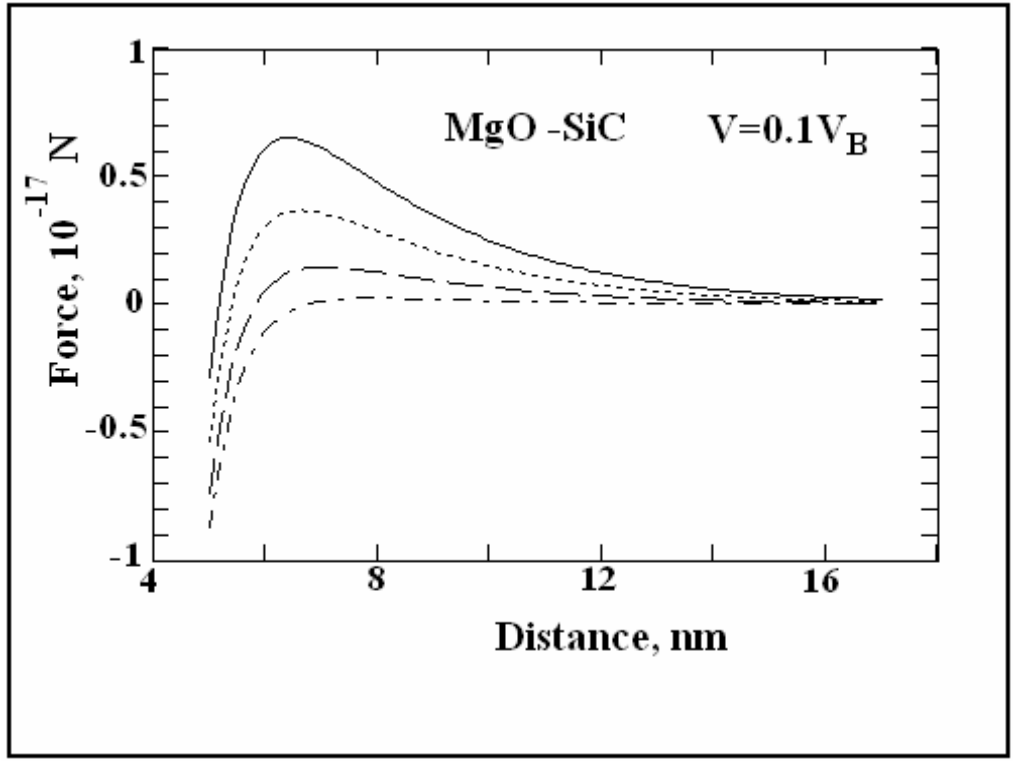

FIGURE 7

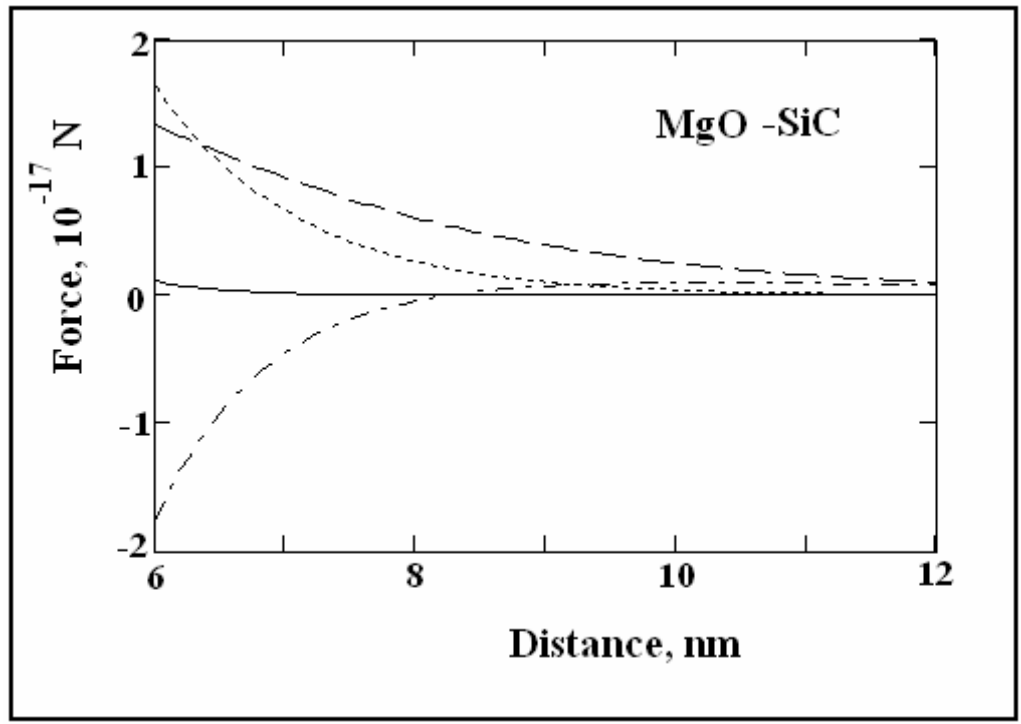